\documentclass{kapproc} 

\usepackage[dvips]{graphicx}

\upperandlowercase

\kluwerbib 

\def\lsim{\mathrel{\rlap{\lower 4pt \hbox{\hskip 1pt $\sim$}}\raise 1pt
\hbox {$<$}}}
\def\gsim{\mathrel{\rlap{\lower 4pt \hbox{\hskip 1pt $\sim$}}\raise 1pt
\hbox {$>$}}}

\newcommand{\eg}{e.g.\ }

\newcommand{\ms}{$M_\odot$}
\newcommand{\Msun}{$M_\odot$}

\newcommand{\kms}{km~s$^{-1}$}

\newcommand{\NaI}{Na~{\sc i}}

\newcommand{\SiII}{Si~{\sc ii}}

\newcommand{\FeII}{Fe~{\sc ii}}

\newcommand{\co}{$^{56}$Co}
\newcommand{\Nifs}{$^{56}$Ni}
\newcommand{\Mej}{$M_{\rm ej}$}
\newcommand{\KE}{$E_{\rm kin}$}

\begin{document}

\articletitle{Hypernovae and Other Black-Hole-Forming Supernovae}


\author{Ken'ichi Nomoto,\altaffilmark{1,2} Keiichi Maeda,\altaffilmark{1}
Paolo A. Mazzali,\altaffilmark{2,3} Hideyuki Umeda,\altaffilmark{1} 
Jinsong Deng,\altaffilmark{1,2} Koichi Iwamoto,\altaffilmark{4} }

\altaffiltext{1}{Department of Astronomy, School of Science, 
University of Tokyo}
\altaffiltext{2}{Research Cenetr for the Early Universe, School of Science, 
University of Tokyo}
\altaffiltext{3}{Osservatorio Astronomico di Trieste, 
Italy}
\altaffiltext{4}{Department of Physics, College of Science and Technology,
Nihon University}

\vspace*{-80mm}
\hspace{-30mm}
\noindent
{\scriptsize 
49 pages, to be published in "Stellar Collapse" (Astrophysics and Space Science; Kluwer) ed. C. L. Fryer (2003)
}
\vspace*{78mm}

\begin{abstract}

During the last few years, a number of exceptional core-collapse
supernovae (SNe) have been discovered. Although their properties are
rather diverse, they have the common feature that at least some of
their basic parameters (kinetic energy of the explosion, mass of the
ejecta, mass of the synthesized $^{56}$Ni), and sometimes all of them,
are larger, sometimes by more than an order of magnitude, than the
values typically found for this type of SNe.  Therefore, these SNe
have been given the collective classification of `Hypernovae'.  The
best known object in this class is SN~1998bw, which owes its fame to
its likely association with the gamma-ray burst (GRB) 980425.  In this
paper, we first describe how the basic parameters of SN~1998bw can be
derived from observations and modeling, and discuss the properties of
other hypernovae individually.  These hypernovae seem to come from
rather massive stars, being more massive than $\sim$ 20 - 25 \ms\ on
the main-sequence, thus forming black holes.  On the other hand, there
are some examples of massive SNe with only a small kinetic energy.  We
suggest that stars with non-rotating black holes are likely to
collapse "quietly" ejecting a small amount of heavy elements (Faint
supernovae).  In contrast, stars with rotating black holes are likely
to give rise to very energetic supernovae (Hypernovae).  We present
distinct nucleosynthesis features of these two types of
"black-hole-forming" supernovae.  Nucleosynthesis in Hypernovae are
characterized by larger abundance ratios (Zn,Co,V,Ti)/Fe and smaller
(Mn,Cr)/Fe.  Nucleosynthesis in Faint supernovae is characterized by a
large amount of fall-back.  We show that the abundance pattern of the
recently discovered most Fe deficient star, HE0107-5240, and other
extremely metal-poor carbon-rich stars are in good accord with those
of black-hole-forming supernovae, but not pair-instability supernovae.
This suggests that black-hole-forming supernovae made important
contributions to the early Galactic (and cosmic) chemical evolution.
Finally we discuss the nature of First (Pop III) Stars.
\end{abstract}


\section{Introduction}

One of the most interesting recent developments in the study of
supernovae (SNe) is the discovery of some very energetic supernovae,
whose kinetic energy (KE) exceeds $10^{52}$\,erg, about 10 times the
KE of normal core-collapse SNe (hereafter $E_{51} = E/10^{51}$\,erg).
The most luminous and powerful of these objects, the Type Ic supernova
(SN~Ic) 1998bw, was probably linked to the gamma-ray burst GRB 980425
(Galama et al. 1998), thus establishing for the first time a 
connection between gamma-ray bursts (GRBs) and the well-studied
phenomenon of core-collapse SNe.  However, SN~1998bw was exceptional
for a SN~Ic: it was as luminous at peak as a SN~Ia, indicating that it
synthesized $\sim 0.5$ \Msun\ of \Nifs, and its KE was estimated at $E
\sim 3 \times 10^{52}$ erg (Iwamoto et al. 1998; Woosley, Eastman, \&
Schmidt 1999).  Because of its large KE, SN~1998bw was called a
``Hypernova (HN)".

\begin{figure}[t]
 \begin{center}
  \includegraphics[width=.7\textwidth]{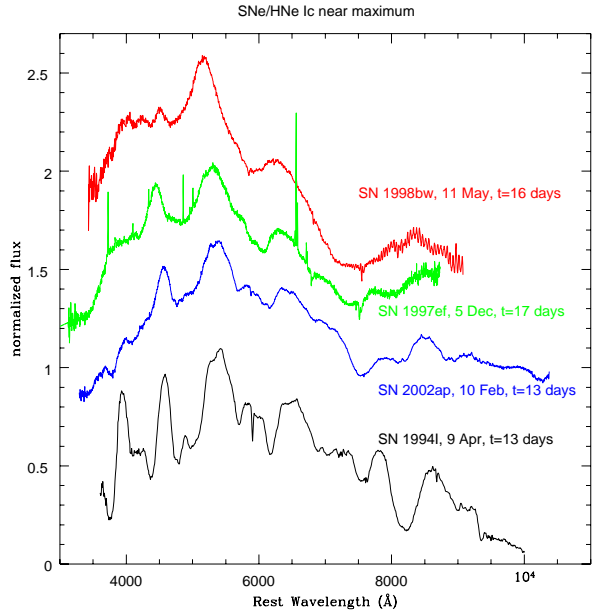}
 \end{center}
 \caption[]{The near-maximum spectra of Type Ic SNe and
 hypernovae: SNe 1998bw, 1997ef, 2002ap, and 1994I
(Mazzali et al. 2002).}  \label{eps1}
\end{figure}

\begin{figure}[t]
 \begin{center}
  \includegraphics[width=.7\textwidth]{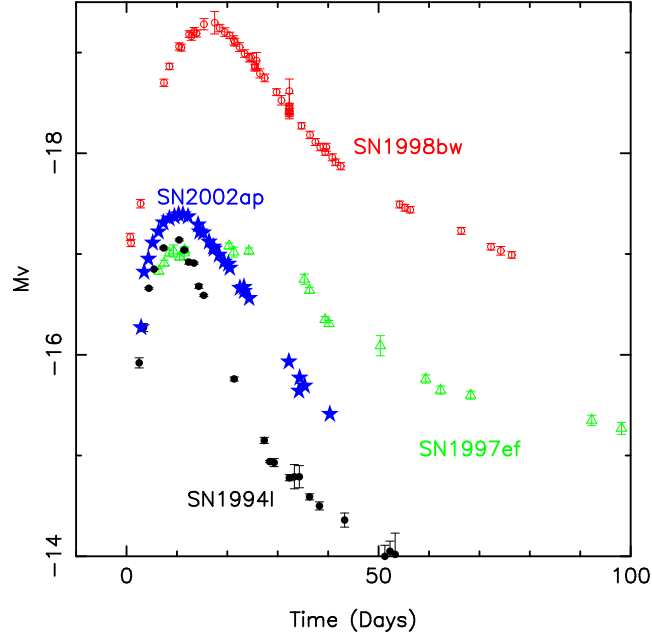}
 \end{center}
 \caption[] {The observed $V$-band light curves of SNe 1998bw ({\em open
circles}), 1997ef ({\em open triangles}), 2002ap ({\em stars}), and
1994I ({\em filled circles}) (Mazzali et al. 2002). }
\label{eps2}
\end{figure}

Subsequently, other ``hypernovae" of Type Ic have been discovered or
recognised, such as SN~1997ef (Iwamoto et al. 2000; Mazzali, Iwamoto
\& Nomoto 2000), SN~1997dq (Matheson et al. 2001), SN~1999as (Knop et
al. 1999), and SN~2002ap (Mazzali et al. 2002).  Although these SNe Ic
did not appear to be associated with GRBs, most recent ``hypernova''
SN 2003dh is clearly associated with GRB 030329 
(Stanek et al. 2003; Hjorth et al. 2003; Kawabata et al. 2003).  
Figures \ref{eps1} and \ref{eps2} show the near-maximum
spectra and the absolute V-light curves of these hypernovae.  These
objects span a wide range of properties, although they all appear to
be highly energetic compared to normal core-collapse SNe.

SN 1999as is the most luminous supernova ever discovered, reaching a
peak magnitude $M_{\rm V} < -21.5$, while the brightness of SN 2002ap
appears to be similar to that of normal core collapse SNe.  The
analysis of these various objects suggests that the KE may be related
to the progenitor's main-sequence mass, which was probably $\gsim 50$
\Msun\ for SN~1999as, $\sim 40$ \Msun\ for SN~1998bw, $\sim 30$ \Msun\
for SN~1997ef, and $\sim 20-25$ \Msun\ for SN~2002ap.  Another
possible hypernovae, although of Type IIn, were SN~1997cy and 1999E,
which was also estimated to have a large mass ($\sim 25$ \Msun;
Germany et al. 2000; Turatto et al. 2000; Rigon et al. 2003).  These
mass estimates place hypernovae at the high-mass end of SN
progenitors, as they are always larger than the mass of the
progenitors of normal core-collapse SNe, which is $\sim 15-20$ \Msun.

In the following sections, we examine the properties of these
hypernovae as derived from optical light curves and spectra and
discuss what may be the discriminating factor for the birth of a
hypernova or the connection with a GRB.  We then focus on
nucleosynthesis in hypernovae, which is quite distinct from the case
of ordinary supernovae, thus making a unique contribution to galactic
chemical evolution.

\section{SN~1998bw and GRB980425}

In the early spectra of SN 1998bw, only few features were present. 
Absorption features were very broad, if present at all, while broad
peaks were present at 4000, 5,000, 6,000 and 8,000 \AA (Galama et
al. 1998).  The absence of any hydrogen lines, of any clear He I
features, and of a strong Si II 6355 line indicated that SN1998bw
should be classified as a Type Ic SN, despite some deviation from
previously known objects of this class, mostly due to the extreme line
width.  SNe Ic are thought
to be the result of core-collapse-induced explosions of C+O stars,
which are massive stars that have lost their H-rich envelope and
almost all of their He shell, through either a wind or binary
interaction (Nomoto et al. 1994).

SN 1998bw is located in a spiral arm of the barred spiral galaxy ESO
184-G82, for which a distance is measured as $\sim38$ Mpc from its
redshift velocity 2550 km s$^{-1}$ (Galama et al. 1998) and a Hubble
constant $H_{0} = 65$ km s$^{-1}$ Mpc$^{-1}$.  Then, the peak absolute
luminosity is estimated to be $\sim 10^{43}$ ergs sec$^{-1}$, which is
about ten times brighter than typical core-collapse SNe (SNe Ib/Ic, or
II).  Assuming that the GRB was at the same distance as SN~1998bw, the
gamma-ray fluence of GRB980425 reported by the BATSE (CGRO) group,
$(4.4 \pm 0.4) \times 10^{-6}$ erg cm$^{-2}$ (Kippen et al. 1998),
corresponds to a burst energy $\sim (7.2 \pm 0.65) \times 10^{47}$
ergs (in gamma-rays). This is four orders of magnitude smaller than
for average GRBs, implying that either GRB980425 might not be a
typical burst or that it may not be related to SN 1998bw.  However,
the recent case of GRB030329/SN2003dh has provided a solid evidence of
connection between the ordinary GRBs and Hypernovae (Stanek et
al. 2003; Hjorth et al. 2003; Kawabata et al. 2003). 
Wang \& Wheeler (1998) studied the correlation between SNe
and GRBs in the literature systematically.

Extensive follow-up observations of SN~1998bw have brought us
invaluable information on this peculiar SN and put rather strong
constraints on corresponding theoretical models.

\subsection{Early Light Curve}

The early light curve of SN 1998bw has been modeled based on
spherically symmetric explosions.  Various investigations reached the
similar conclusion that the early light curve can be successfully
reproduced by the explosion of a massive C+O star with a kinetic
energy more than ten times larger than a canonical SN explosion
(Iwamoto et al. 1998; Woolsey et al. 1999).  Figure \ref{eps5}
compares the light curve of SN 1998bw with those of other SNe Ic and
their model light curves (Table 1).

SN 1998bw showed a very early rise. It had a luminosity of $\sim
10^{42}$ erg s$^{-1}$ already at day 1.  This rapid rise requires the 
presence of $^{56}$Ni near the surface. Since spherically symmetric
explosion models produce $^{56}$Ni in deep inner layers of the ejecta,
they can only reproduce this behaviour if extensive mixing is postulated 
to have occurred, dredging $^{56}$Ni up to outer layers
(Iwamoto et al. 1998; Nakamura et al. 2001a).

The light curve of SN 1998bw reaches the peak on about day 16, with
the peak absolute magnitude being comparable to that of normal SNe Ia.
We obtain a qualitative guess of model parameters such as ejecta mass
$M_{\rm ej}$, explosion kinetic energy $E_{\rm k} = E_{51} \times
10^{51}$ erg, and the $^{56}$Ni mass $M_{\rm Ni}$ by employing an
analytic solution by Arnett (1982). At early times, the bolometric
luminosity $L_{\rm bol}$ of a ``compact'' SN that is powered by
radioactive decay of $^{56}$Ni is written as

\begin{equation}
L_{\rm bol} = \frac{\epsilon_{\rm Ni} M_{\rm Ni}}{\tau_{\rm Ni}} \Lambda(x,y),
\end{equation}

\noindent where $\tau_{\rm Ni}= 8.8$ days and
$\epsilon_{\rm Ni}= 2.96 \times 10^{16}$ erg g$^{-1}$ are the decay
time of $^{56}$Ni and the energy deposited per gram of $^{56}$Ni,
respectively, and the function $\Lambda$ is given by

\begin{equation}
\Lambda(x,y) = \exp(-x^{2})\int_{0}^{x} \exp(-2zy+z^{2}) 2z dz.
\end{equation}

It is assumed that $^{56}$Ni is distributed homogeneously and that
$\gamma$-rays are all trapped in the ejecta. This latter assumption is
correct at the early phases we are considering.  The dimensionless
variable $x = t/\tau_{c}$, where $t$ is the elapsed time and
$\tau_{c}$ is the characteristic time of the light curve $\tau_{c} =
(2 \tau_{h} \tau_{d})^{1/2}$, where $\tau_{h}$ and $\tau_{d}$ are the
hydrodynamical time scale and the diffusion time scale of optical
photons through the ejecta, respectively.  The variable $y$ is defined
as $y = \tau_{c}/(2 \tau_{\rm Ni})$.  For the principal mode of
diffusion (Arnett 1982; Pinto \& Eastman 2000), $\tau_{c}$ turns out
to be

\begin{equation}
\tau_{c} \sim 8~{\rm days} \left(\frac{\kappa}{0.05}\right)^{1/2}
M_{\rm ej, \odot}^{3/4} E_{\rm k, 51}^{-1/4}
\end{equation}

\noindent
where $\kappa$ is the effective opacity.  Hereafter, the symbol $M_{\rm x,
\odot}$ denotes $M_{\rm x}/M_{\odot}$.  The ejecta are
assumed to be a sphere of constant density expanding homologously, and
thus $M_{\rm ej}$ and $E_{\rm k} = E_{\rm k, 51} \times 10^{51}$ erg
are given by $M_{\rm ej} = (4 \pi/3)R^{3}\rho $, $E_{\rm k} \sim
(3/5) (1/2) M_{\rm ej} v_{s}^{2}$, where $R$, $\rho$, and $v_{s}$
are the radius, density, and surface velocity of the ejecta, respectively.

\begin{figure}[t]
 \begin{center}
  \includegraphics[width=.5\textwidth]{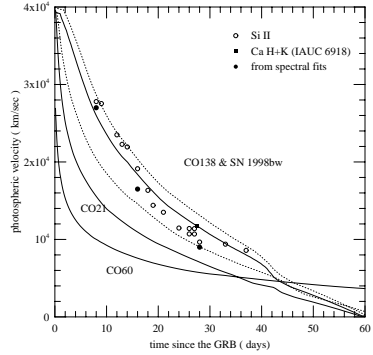}
 \end{center}
\vskip -2cm
\caption[]{Photospheric velocities of SN 1998bw.  The photospheric
 velocities obtained by spectral modeling (filled circles) and the
 observed velocities of Si II 6347 \AA, 6371 \AA \ lines measured at the
 absorption cores (open circles: Patat et al. 2001), and that of the Ca II
 H+K doublet derived from the spectrum of 23 May 1998 (filled square)
 are compared with a series of C+O star explosion models
 (Iwamoto et al. 1998).}  \label{eps3}

\end{figure}

\begin{table}
\begin{center}
\renewcommand{\arraystretch}{1.4}

\setlength\tabcolsep{5pt}
\begin{tabular}{lllllll}
\hline\noalign{\smallskip}
& $M_{\rm ms}$
& $M_{\rm C+O}$
& $M_{\rm ej}$
& $^{56}$Ni mass
& $M_{\rm cut}$
& $E_{\rm K}$
\\
Model
& {\small($M_\odot$)}
& {\small($M_\odot$)}
& {\small($M_\odot$)}
& {\small($M_\odot$)}
& {\small($M_\odot$)}
& {\small (10$^{51}$ erg)} \\
\noalign{\smallskip}
\hline
\noalign{\smallskip}
CO138E50 & $\sim$ 40 & 13.8 & 10   & 0.4  &  4   & 50 \\
CO138E30 & $\sim$ 40 & 13.8 & 10.5 & 0.4  &  3.5 & 30 \\
CO138E7  & $\sim$ 40 & 13.8 & 11.5 & 0.4  &  2.5 &  7 \\
CO138E1  & $\sim$ 40 & 13.8 & 12   & 0.4  &  2   &  1 \\
CO100    & $\sim$ 30 & 10.0 & 7.6  & 0.15 &  2.4 & 10 \\
CO60     & $\sim$ 25 &  6.0 & 4.4  & 0.15 &  1.4 &  1 \\
CO21     & $\sim$ 15 &  2.1 & 0.9  & 0.07 &  1.2 &  1 \\
\hline
\end{tabular}
\end{center}
\caption{Models and their parameters for SNe Ic. CO138E1 is an
 ordinary SN Ic model, in which a C+O star of $M_{\rm CO}=
 13.8 M_\odot$ (which is the core of a $\sim 40 M_\odot$ main-sequence
 star) explodes with $E_{\rm K} = 1 \times 10^{51}$ ergs and $M_{\rm
 ej} = M_{\rm CO}-M_{\rm cut} \simeq 12 M_\odot$.  $M_{\rm cut}$ (= 2
 $M_\odot$ in this case) denotes the mass cut, which corresponds to
 the mass of the compact star remnant, either a neutron star or a
 black hole. CO138E50, CO138E30, and CO138E7 are hypernova models, in
 which the progenitor C+O star is the same as CO138E1 but explodes
 with different energies.  The mass cut is chosen so that the ejected
 mass of $^{56}$Ni is the value required to explain the observed peak
 brightness of SN~1998bw.  }  \label{Tab1.1a}
\end{table}

Differentiating $\Lambda$ with respect to $x$, it is found that
$\Lambda$ has a maximum $\Lambda_{\rm max} = \exp(-2xy)$.  As given
in Table 1 of Arnett (1982), the maximum occurs at 
$x_{\rm max} y = t_{\rm max}/(2 \tau_{\rm Ni}) \sim 0.42+0.48y$ and
$\Lambda_{\rm max} \sim 0.165/y$ (Arnett 2001).
For SN 1998bw, $t_{\rm max} \sim 16~{\rm days}$ corresponds to 
$y \sim 1$ and $\Lambda_{\rm max} \sim 0.165$, thus $L_{\rm max} \sim
1.3 \times 10^{43} M_{\rm Ni, \odot} {\rm erg~s}^{-1}$. Given the fact
that the $^{56}$Co contribution doubles this luminosity at day 
$\sim 16$, the mass of $^{56}$Ni is approximately given by
\begin{eqnarray}
M_{\rm Ni, \odot} \sim 0.38 L_{\rm max, 43}
\end{eqnarray}
\noindent
where $L_{\rm max, 43}$ is the peak luminosity in units of $10^{43}$
erg s$^{-1}$.  The observed peak luminosity of SN 1998bw, $L_{\rm
max,43} \sim 1 \times (d/37.8 {\rm Mpc})^{2}$ with $d$ being the
distance, translates into an estimated mass of $^{56}$Ni $M_{\rm Ni,
\odot} \sim 0.4(d/37.8 {\rm Mpc})^{2}$, which is much larger than that
of typical core-collapse SNe $M_{\rm Ni, \odot} \lsim 0.1$.

A constraint for $M_{\rm ej, \odot}$ and $E_{\rm k, 51}$ can be
obtained from $y \sim 1$ with the use of equation (3) such that

\begin{eqnarray}
M_{\rm ej, \odot}^{3}/E_{\rm k, 51} \sim 23
\left(\frac{\kappa}{0.05}\right)^{-2}
\end{eqnarray}

Another equation is necessary to resolve the degeneracy of masses and
energies in this equation.  One useful quantity to use is the
evolution of the photospheric velocity, as this has a different
dependence on the parameters $M_{\rm ej, \odot}$ and $E_{\rm k, 51}$.
The photospheric velocity can either be obtained from observations,
determining the approximate velocity of absorption lines, or it can be
computed in the explosion models. An equivalent but more
quantitative approach is to compare the observed and synthetic spectra.

In Figure \ref{eps3}, the observed line velocities and photospheric
velocities of SN 1998bw are compared with calculated photospheric
velocities for different models.  The radius of the photosphere $r_{\rm
ph}$ is defined by $ \int_{r_{\rm ph}}^{\infty} \kappa \rho dr = 2/3$.  
For the constant density sphere model used in analytic light curves, 
the photosphere is located at a fraction $\sim 80\pi E_{\rm k}/(27 
\kappa M_{\rm ej}^{2}) t^{2}$ of the ejecta radius from the surface.  
Then, the photospheric velocity $v_{\rm ph}$ at early phases is given by

\begin{eqnarray}
v_{\rm ph} \sim v_{s} 
\left(1-\frac{80\pi E_{\rm k}}{27 \kappa M_{\rm ej}^{2}} t^{2}\right) 
= \left(\frac{10E_{\rm k}}{3 M_{\rm ej}}\right)^{1/2}
\left(1-\frac{80\pi E_{\rm k}}{27 \kappa M_{\rm ej}^{2}} t^{2}\right).
\end{eqnarray}

Using the fact that on day 8 $v_{\rm ph} \sim 27,000$ \kms, we obtain
from equation (6) another constraint:

\begin{eqnarray}
E_{\rm k, 51}/M_{\rm ej, \odot} \sim 4.3 \nonumber
\end{eqnarray}

\noindent
which, combined with equation (5), results in

\begin{eqnarray}
M_{\rm ej, \odot} \sim 10, \hskip 1cm E_{\rm k, 51} \sim 43.
\end{eqnarray}

This very large energy, exceeding $10^{52}$ erg, led us to refer to SN
1998bw as a ``Hypernova (HN)'', an exceptional class of energetic SN
explosion with kinetic energy $E_{\rm k, 51} \gsim 10$ (Iwamoto et
al. 1998; Woosley et al. 1999).  The above estimate provides model
parameters in good agreement with those given by detailed light curve
models (Iwamoto et al. 1998; Woosley et al. 1999; Nakamura et al. 2001a).
Nakamura et al. (2001a) selected a model (CO138E50) with $M_{\rm ej,
\odot} = 10, E_{\rm k, 51} = 50, M_{\rm Ni, \odot} = 0.4$ as the best
model to reproduce the light curve and photospheric velocities using a
distance $ 37.8 $Mpc and an extinction $A_{\rm V} = 0.05$ for SN1998bw
(see Table 1 and Fig.~\ref{eps5}).

\begin{figure}[t]
 \begin{center}
  \includegraphics[width=.7\textwidth]{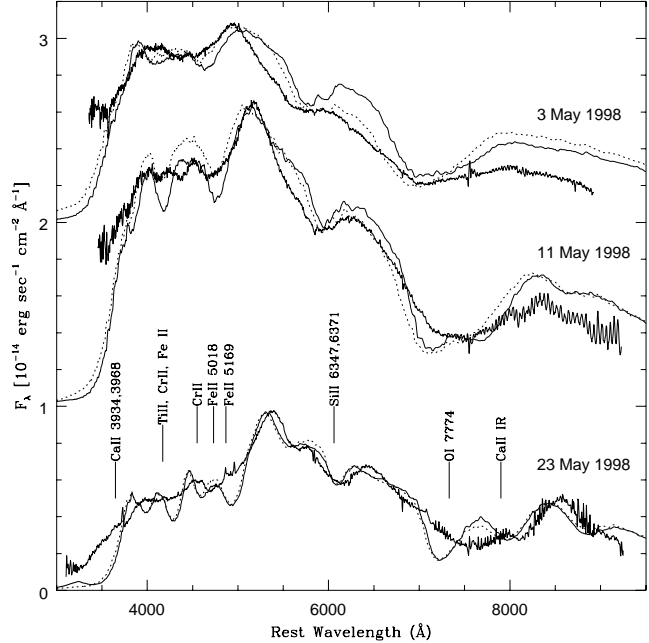}
 \end{center}
 \caption[]{The observed spectra of SN 1998bw at three epochs: 3, 11,
 and 23 of May 1998 (day 8, 16, and 28, respectively: Patat et
 al. 2001) are compared with synthetic spectra (dashed lines) using
 model CO138E50 (Nakamura et al. 2001a).  A distance modulus $\mu =
 32.89$ mag and $A_{\rm V}= 0.05$ are assumed, which corresponds to a
 distance of 37.8 Mpc with $H_{0} = 65$ km s$^{-1}$ Mpc$^{-1}$.  The
 observed featureless spectra are the result of blending of many metal
 lines having large velocities and large velocity spreads. The
 apparent emission peaks are actually low opacity regions of the
 spectra in which photons can escape. } \label{eps4}
\end{figure}

\subsection{Early Spectra}

A more quantitative constraint on these basic model parameters and
crucial diagnosis of chemical compositions of the ejecta can be
obtained by detailed spectral modeling (Iwamoto et al. 1998; Branch
2001; Nomoto et al. 2001ab; Nakamura et al. 2001a; Mazzali et
al. 2001).

Figure~\ref{eps4} shows early spectra of SN 1998bw for three epochs
(Nakamura et al. 2001a; Patat et al. 2001). The spectra are dominated
by broad absorption features. These line features are also seen in
other SNe Ic.  For SN 1998bw, however, they are exceptionally broad
and blueshifted.  Stathakis et al. (2000) found that absorption line
minima are shifted 10-50 per cent blueward at day 15 in comparison
with ordinary SNe Ic.  These features shift significantly to the red
over the three weeks covered by the spectra shown, indicating how the
photosphere is receding to inner, lower velocity parts of the
ejecta. The bluest of these features is likely due to Fe II lines,
while the feature near 6000\AA\ is dominated by Si II and the redmost
one is a blend of O I and Ca II. That the O I and Ca I lines merge
into a single broad absorption is very unusual for any SN, and it
indicates that the ejecta velocities are very large (the line
separation is $\sim 30000$km/s).

In Fig.~\ref{eps4}, the observed spectra are compared with synthetic
spectra computed with a hypernova model CO138E50 (Nakamura et
al. 2001a).  A distance modulus of $\mu = 32.89$~mag and an extinction
$A_V = 0.05$ are adopted.  The assumption of low reddening is
supported by the upper limit of 0.1~\AA\ in the equivalent width of
Na~{\sc i}~D line obtained from high-resolution spectra (Patat et
al. 2001).

The synthetic spectra of CO138E50 are in good agreement with the
observed spectra.  The Si II feature near 6,000\AA\ and, in
particular, the OI+CaII feature between 7,000 and 8,000\AA, are as
broad as the observations.  Nevertheless, the blue sides of those
absorptions are still too narrow, indicating that the new model
CO138E50 may not contain enough mass in the high velocity part.
Nakamura et al.  (2001a) tried a model CO138E50 with its density
structure in the envelope made shallower artificially and examined the
effect on the synthetic spectra.  They found that a reduction of
density gradient from $\rho \propto r^{-8}$ to $\rho \propto r^{-6}$
at $v > 30,000$ km s$^{-1}$ leads to a significant increase of mass in
higher velocity regions, making strong absorption features at $v \sim
60,000$ km s$^{-1}$.  Similar conclusions were reached by Branch
(2001), who presented parameterized synthetic spectra for the early
phases of SN~1998bw.

\subsection{Late-time Light Curves and Spectra}

\begin{figure}[t]
 \begin{center}
\hskip -2cm
  \includegraphics[width=.7\textwidth]{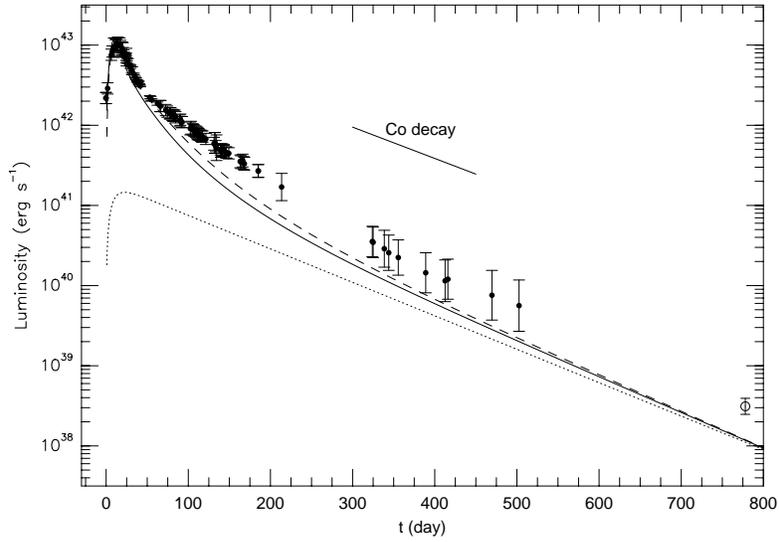}
 \end{center}
 \caption[]{Light curves of higher-energy models (Nakamura et al. 2001a).
The light curves of CO138E50 ($E_{\rm K} = 5 \times 10^{52}$ erg; solid
line) and CO138E30 ($E_{\rm K} = 3 \times 10^{52}$ erg; long-dashed
line) compared with the bolometric light curve of SN1998bw
(Patat et al. 2001).  A distance modulus of $\mu = 32.89$ mag and $A_V =
0.05$ are adopted.  The dotted line indicates the energy deposited by
positrons for CO138E50.  The HST observation at day 778 (Fynbo 2000)
is shown by assuming negligible bolometric correction.  } \label{eps5}
\end{figure}

The dominant contributions to the light curve change as the SN evovles. The
light curve is initially powered by $\gamma$-rays from $^{56}$Co, then by
positrons of the $^{56}$Co decay, and finally by $\gamma$-rays from
$^{57}$Co and positrons from $^{44}$Ti.

The optical depth of the ejecta to the $\gamma$-ray photons produced by
the $^{56}$Co decay, $\tau_{\gamma}$, is given by

\begin{eqnarray}
\tau_{\gamma} = \kappa_{\gamma} \rho R
= \frac{9 \kappa_{\gamma} M_{\rm ej}^{2}}{40 \pi E_{\rm k}} t^{-2}
\sim
0.11 \left(\frac{\kappa_{\gamma}}{0.03}\right)
\frac{M_{\rm ej, \odot}^{2}}{E_{\rm k, 51}}
\left(\frac{t}{100 \ {\rm days}}\right)^{-2}
\end{eqnarray}

\noindent where $\kappa_{\gamma}$ is the effective $\gamma$-ray
opacity for the $^{56}$Co gamma-ray lines.  In model CO138E50 for SN
1998bw, the ejecta become optically thin in $\gamma$-rays
($\tau_{\gamma} < 1$) at around day 50. Using a deposition fraction
$f_{\rm dep, \gamma} = 0.64 \tau_{\gamma}$ for $\tau_{\gamma} <
0.25$ (Colgate , Petschek, \& Kriese 1980), we find

\begin{eqnarray} 
f_{\rm dep, \gamma} \sim 0.073 
\left(\frac{\kappa_{\gamma}}{0.03}\right) 
\frac{M_{\rm ej, \odot}^{2}}{E_{\rm k, 51}} 
\left(\frac{t}{100 \ {\rm days}}\right)^{-2}. 
\end{eqnarray}

Assuming that the energy deposited by $\gamma$-rays is thermalized and 
subsequently radiated in optical/IR wavelengths, the bolometric
luminosity $L_{\rm bol}$ is given by

\begin{eqnarray}
L_{\rm bol}
\sim f_{\rm dep, \gamma} \epsilon_{\rm decay} M_{\rm Ni},  \nonumber
\end{eqnarray}

\noindent
where $\epsilon_{\rm decay}$ is the energy available per gram of
$^{56}$Co per second

\begin{eqnarray}
\epsilon_{\rm decay}
= \frac{\epsilon_{\rm Ni}}{\tau_{\rm Ni}} \exp
\left(-\frac{t}{\tau_{\rm Ni}}\right)
+ \frac{\epsilon_{\rm Co}}{\tau_{\rm Co}-\tau_{\rm Ni}}
\left[ \exp \left(-\frac{t}{\tau_{\rm Co}}\right)-\exp \left(-\frac{t}{\tau_{\rm
Ni}}\right) \right],
\end{eqnarray}

\noindent
where $\epsilon_{\rm Co} = 6.3 \times 10^{16}$ erg g$^{-1}$ and
$\tau_{\rm Co} = 111.3$ days are the energy deposited per gram of
$^{56}$Co and its decay time, respectively.  Since at late times
most of the energy deposition comes from the $^{56}$Co decay, we have

\begin{eqnarray}
& & L_{\rm bol} \sim \frac{f_{\rm dep, \gamma} \epsilon_{\rm Co} M_{\rm Ni}}
{\tau_{\rm Co}-\tau_{\rm Ni}} \exp \left(-\frac{t}{\tau_{\rm Co}}\right)
\nonumber
\\
& \sim & 1.6 \times 10^{41} {\rm erg \ s}^{-1}
\left(\frac{\kappa_{\gamma}}{0.03}\right)
\left(\frac{M_{\rm Ni, \odot}}{0.4}\right) 
\frac{M_{\rm ej, \odot}^{2}}{E_{\rm k, 51}}
\left(\frac{t}{111 {\rm d}}\right)^{-2} \hspace{-0.4cm}
\exp \left(-\frac{t}{111 {\rm d}}\right). \hspace{0.6cm}
\end{eqnarray}

\noindent
This luminosity depends on $M_{\rm ej}$ and $E_{\rm k}$ as well as on
$M_{\rm Ni}$. Thus we can use late-time light curves to distinguish
between different models that produce similar early light curves.
For CO138E50, equation (11) gives $L_{\rm bol} \sim 2 \times
10^{40}$ erg s$^{-1}$ at day 400, which is in agreement with the
observations of SN 1998bw (Fig.~\ref{eps5}).

Figure~\ref{eps5} compares light curves calculated for hypernova models
CO138E50 (solid line) and CO138E30 (long-dashed line) with the
observed light curve of SN 1998bw (filled circles).  The light curve
of CO138E50 is consistent with the observations until day 50 but
declines at a faster rate afterwards. The light curve of CO138E30
shows a slower decline, in better agreement with SN1998bw, but it
still declines too rapidly (Nakamura et al. 2001a; McKenzie \&
Schaefer 1999; Patat et al. 2001). All of these models (see Table 1)
originate from a 13.8$M_\odot$ CO core, which is formed in a $\sim
40M_\odot$ star. The explosion may have given rise to a Black Hole.

The early light curve of SN 1998bw is well reproduced by higher-energy
models, but they deviate from the observations at late times.  On the
other hand, the late-time light curve is more easily reproduced by a
lower-energy model with a smaller $^{56}$Ni mass (model CO138E7),
which is however too dim at early phases (Nakamura et al. 2001a).
This difficulty may be overcome if the ejecta have multiple components
with different characteristic velocities, suggesting that either the
density distribution or the $^{56}$Ni distribution in the ejecta or
both are not spherically symmetric.  A similar situation occurrs for 
SN~1997ef (Iwamoto et al. 2000; Mazzali et al. 2000) and SN~2002ap
(Maeda et al. 2003).

As seen in Fig.\ref{eps5}, the light curve of SN~1998bw seems to
flatten after about day 400.  The model light curves do not really
follow this behavior. At $ t \gsim 400$ days, the $\gamma$-ray
deposition fraction $f_{\rm dep, \gamma}$ decreases to below 1\% in
model CO138E50.  However, about 3.5\% of the decay energy of $^{56}$Co
is carried by positrons (e.g., Axelrod 1980).  These are effectively
trapped in the ejecta because of the postulated weak magnetic fields
(e.g. Colgate \& Petschek 1979).  Therefore, the energy deposition
from positrons makes the dominant contribution to the light curve at
$t \gsim 400$ days (dotted line in Fig. \ref{eps5}).

\begin{figure}[t]
 \begin{center}
  \includegraphics[width=.5\textwidth]{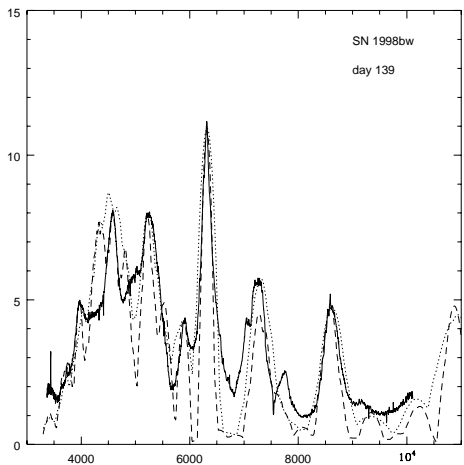}
 \end{center}
 \caption[]{ A nebular spectrum of SN~1998bw on 12 Sept 1998 (day 139)
is compared to synthetic spectra obtained with a NLTE nebular model
based on the deposition of gamma-rays from $^{56}$Co decay in a nebula
of uniform density.  Two models were computed.  In one model (dotted
line) the broad FeII lines near 5300\AA \ is well reproduced.  The
derived $^{56}$Ni mass is 0.65 $M_\odot$, and the outer nebular velocity
is 11,000\kms, and the O mass is 3.5$M_\odot$. The average electron
density in the nebula is log $n_e$ = 7.47 cm$^{-3}$.  In the other model
(dashed line), only the narrow OI 6300\AA\ emission line is well
reproduced. This model has smaller $^{56}$Ni mass (0.35 $M_\odot$) and O
mass (2.1 $M_\odot$), and an outer velocity of 7500\kms.  The density is
similar to that of the 'broad-lined' model.  The filling factor used is
0.1 for both models (Nomoto et al. 2001ab; Mazzali et al. 2001).}

\label{eps7}
\end{figure}

The luminosity provided by the positrons is given as

\begin{eqnarray}
L_{\rm bol, e^+} = 2 \times 10^{41}
\left(\frac{M_{\rm Ni, \odot}}{0.4}\right)
\exp \left(-\frac{t}{\rm 111 d}\right)
\end{eqnarray}

\noindent where the positrons are assumed to be completely trapped in
the ejecta.  Therefore, if the observed tail follows the
positron-powered, exponentially declining light curve, the $^{56}$Co
mass can be determined directly.  However, the light curve of
SN~1998bw after day 400 is still steeper than the positron-dominated
light curve.  The light curve of SN~1998bw showed a further flattening
at around day 800.  The latest observed point in Figure~\ref{eps5} is
the HST observation on June 11, 2000 (day 778) (Fynbo 2000).

The observed magnitude (V = 25.41 $\pm$ 0.25) is consistent with the
prediction of CO138E7, but brighter than CO138E50 (Fig.\ref{eps5}). 

One possibility is that, as the density decreases, the recombination
time scale becomes longer than the decay time and ionization
freeze-out makes the bolometric light curve even flatter (Fransson \&
Kozma 1993).  Another possible source of the excess luminosity is the
emission of radiation due to the interaction of the ejecta with a
circumstellar medium (CSM) (Sollerman et al. 2000).  Finally, the
flattening may be due to a contribution from an underlying star
cluster (Fynbo 2000) rather than to the SN itself.

Late time spectra provide a wealth of information on the elemental
abundances and their distributions in velocity space.  SN 1998bw seems
to have entered the nebular phase between day 65 and 115 (Mazzali et
al. 2001; Patat et al. 2001).  The spectroscopic features at late
times are very similar to those of SNe Ic as shown in Figure~\ref{eps7}.  
Dominant emission features
include Mg I$\lambda$4571, an Fe II blend around 5,200 \AA, the O
I$\lambda\lambda$6300,6364 doublet, a feature around 7,200 \AA
(identified as Ca II and C II by Mazzali et al. 2001), and the Ca II
IR triplet.  What is different from ordinary SNe Ic, in particular, is
the broadness of the line features.  Patat et al. (2001) estimated the
expansion velocity of the Mg I emitting region and found a value 9,800
$\pm$ 500 \kms on day 201. Stathakis et al. (2000) also found emission
features 45 per cent broader than ordinary SNe Ic for day 94.

Mazzali et al. (2001) showed that the late time spectra of SN 1998bw
(Patat et al. 2001) contain both broad and relatively narrow lines.
They interpreted the absence of Fe III nebular lines as a sign of 
clumpiness of the ejecta. Interestingly, some Fe lines are found to be
broader than O lines, which is the opposite as to what spherically
symmetric models predict. This may imply that the explosion was
aspherical (Fig.~\ref{eps7}; Mazzali et al. 2001).

\subsection{Aspherical Models}

The result that the explosion energy of SN 1998bw was extremely large
is based on the analysis of the light curve and spectra and on the
assumption that the explosion was spherically symmetric.  There is a
possible alternative explanation for the large peak luminosity of SN
1998bw. If the explosion was aspherical, the estimate of the explosion
energy might be smaller.  There is growing evidence that light from
SNe Ib/c is weakly polarized ($\sim 1$\%), implying that these
explosions might be somewhat aspherical.  A similar degree of
polarization was observed in SN~1998bw at early photospheric phases
(Kay et al. 1998; Iwamoto et al. 1998; Patat et al. 2001).

As discussed earlier, it is difficult to reproduce the entire light
curve of SN 1998bw consistently using spherically symmetric models,
indicating that there exists some degree of asphericity in the
explosion of SN 1998bw.  An analysis of the late-time light curve
suggested a $^{56}$Ni mass $ M_{\rm Ni, \odot}> 0.4$ (Sollerman et al. 
2000), which is close to the values obtained with spherically
symmetric models for early light curves. Thus, the possible effects of
asymmetry in SN 1998bw seem to be only moderate.

\begin{figure}[t]
 \begin{center}
  \includegraphics[width=.7\textwidth]{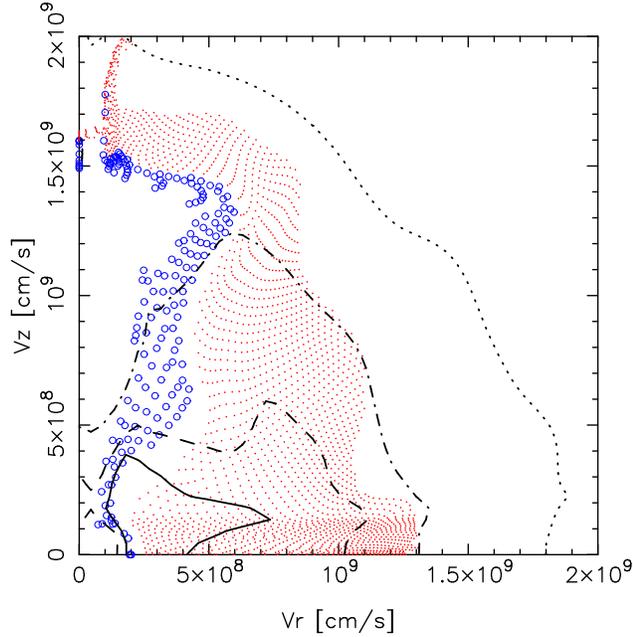}
 \end{center}
 \caption[]{Distribution of $^{56}$Ni(open circles) and $^{16}$O(dots)
at the homologous expansion phase in an aspherical explosion model of SN
1998bw (Maeda et al. 2002). Parameters are $E_{\rm exp} = 10^{52}$ erg,
$v_z/v_r = 8$. Open circles and dots are test particles of $^{56}$Ni
and $^{16}$O, respectively, indicating local volumes in which mass
fractions of these elements are greater than 0.1. Lines are density
contours of 0.5(solid), 0.3(dashed), 0.1(dash-dotted), and 0.01(dotted)
of the maximum density.}  \label{eps10}
\end{figure}

In an asymmetric explosion, nucleosynthesis should also depart from
spherical symmetry.  Maeda et al. (2002) calculated nucleosynthesis in
aspherical explosion models for SN 1998bw with a 2D hydrodynamical
code and a detailed nuclear reaction network.  They used the
progenitor model CO138 (Iwamoto et al. 1998; Nakamura et al. 2001a)
and assumed aspherical initial velocity profiles.  Figure \ref{eps10}
shows the composition in the ejecta at the homologous expansion phase
for $E_{\rm exp} = 10^{52}$ erg and an initial 
axial-to-radial velocity ratio
$v_z/v_r = 8$.  In this model, $^{56}$Ni is synthesized preferentially
along the polar axis, where the shock is stronger, while a lot of
unburned material, dominated by O, is left at low velocity in the
equatorial region, where burning is much less efficient.

Maeda et al. (2002) found that the nebular line profiles in SN 1998bw
can be reproduced by such an aspherical model if the explosion is
observed at an angle of about 15 degrees from the polar axis.  At such
an angle, one might expect that the GRB is weaker than it would be if
observed along the jet axis.  The actual aspect ratio of the ejecta is
much smaller than 8:1, however, as the jet expands laterally, and this
may be consistent with the observed polarization.

Such a highly aspherical explosion could occur in the collapse of a 
rotating stellar core that forms a system consisting of a rotating
black hole and an accretion torus around it (MacFadyen \& Woosley
1999).  Thermal neutrinos from the torus release a large amount of
energy as electron-positron pairs. If the black hole is accompanied by
a strong magnetic field, rotational energy may be extracted from the
black hole via the Blandford-Znajek mechanism (Blandford \& Znajek
1977).  Jet formation and propagation after energy deposition by the
above processes are studied using hydrodynamical simulations
(MacFadyen, Woosley, \& Heger 2001; Aloy et al. 2000). MacFadyen et
al. (2001) suggested that SN 1998bw may be a case in which a black
hole was produced by 'fall back' and the resulting jet was less
collimated.

The aspherical SN explosion could also be induced in the neutron star
formation, e.g., by a strong magnetic field (Nakamura 1998; Wheeler
2001) and convection driven by neutrino heating (e.g., Janka \& M\"uller 
1994; Fryer \& Warren 2002). Shimizu et al. (2001) pointed out that
anisotropy in the neutrino emission would increase the net energy gain
by neutrino heating, which leads to a larger explosion energy than in
spherically symmetric models.

\section{SN~1997ef}\label{sec:97ef}

This SN was an immediate precursor of SN~1998bw. Its broad-lined
spectrum defied interpretation in the context of standard-energy
explosion models (Iwamoto et al. 2000). The realization with SN~1998bw
that energies much larger than the supposedly standard value
$10^{51}$erg were possible led to the reinterpretation of SN~1997ef as
a hypernova, although of smaller energy than SN~1998bw.

\subsection{Light Curve }

\begin{figure}[t]
 \begin{center}
  \includegraphics[width=.45\textwidth]{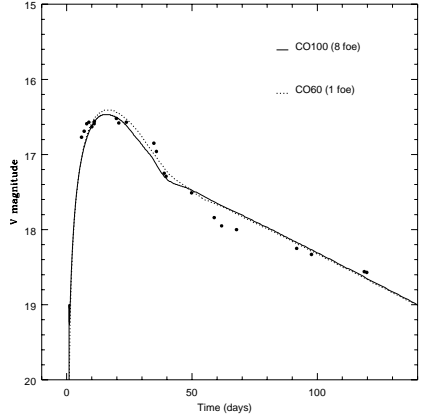}
 \end{center}
 \caption[] {Calculated Visual light curves of CO60 and CO100 compared
with that of SN 1997ef.}
\label{fig:lc97ef}
\end{figure}

\begin{figure}[t]
 \begin{center}
  \includegraphics[width=.6\textwidth]{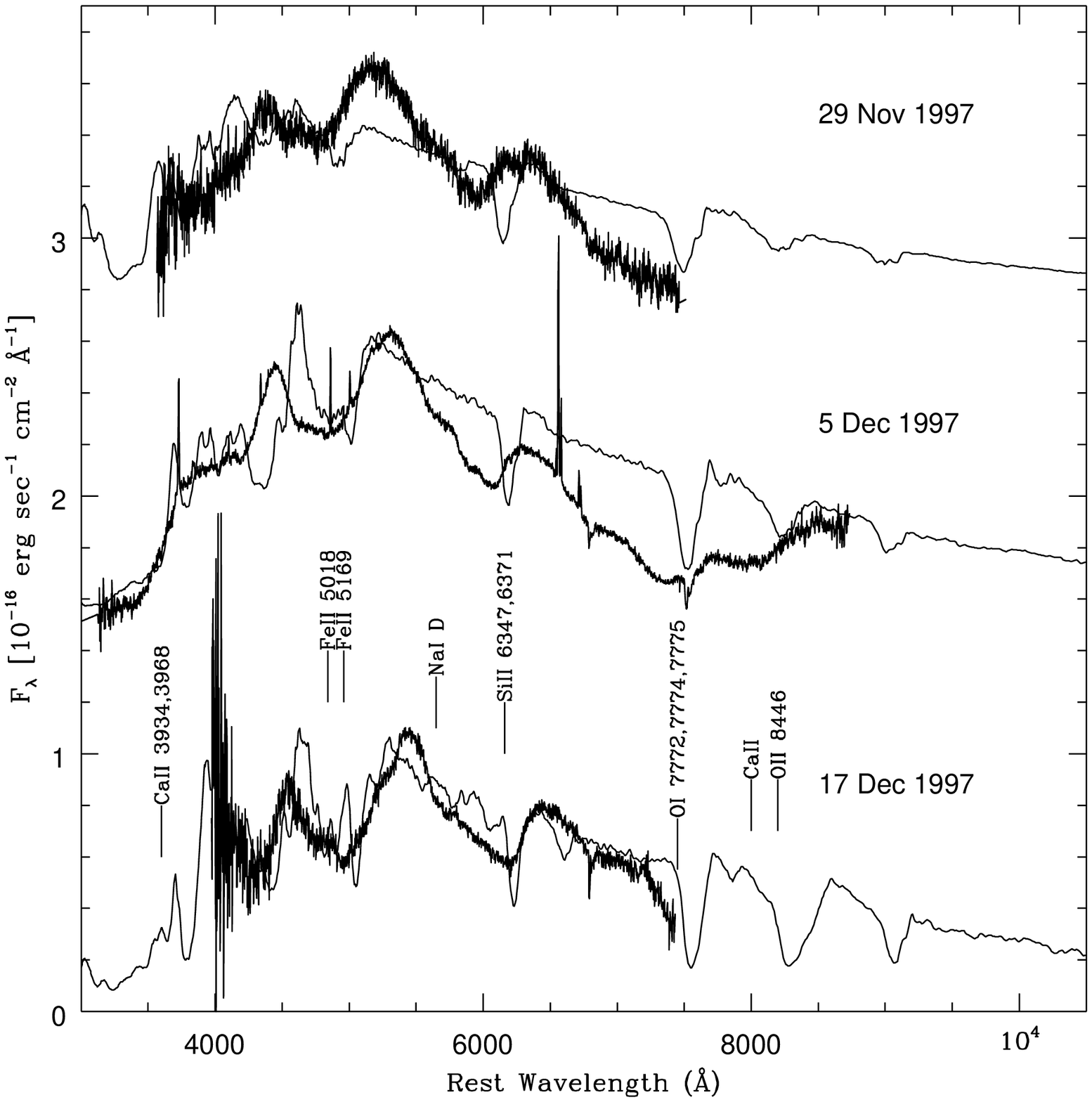}
\vskip 0.2cm
  \includegraphics[width=.6\textwidth]{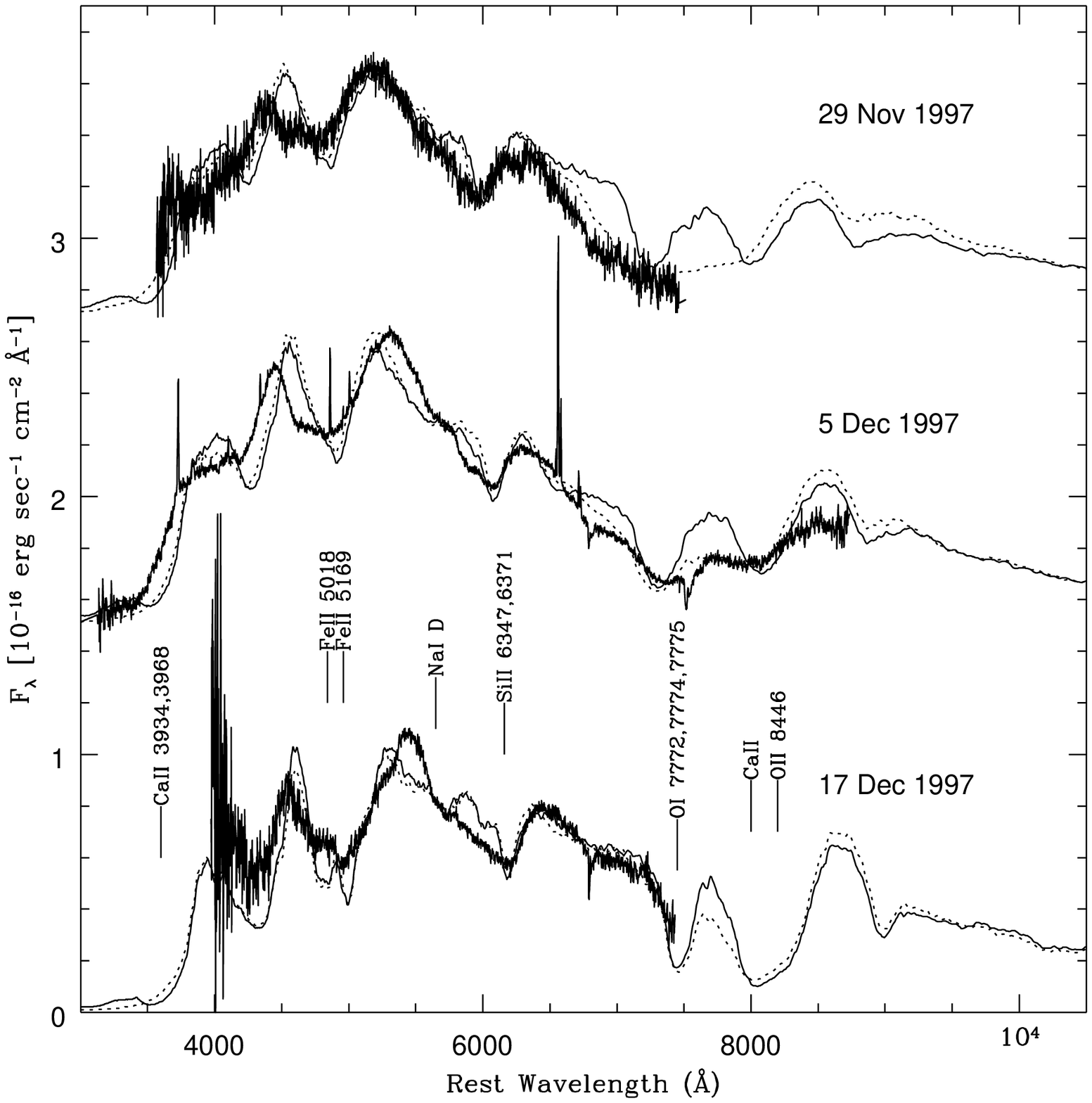}
\end{center}
 \caption[] {
Observed spectra of SN~1997ef (bold lines) and synthetic spectra
computed using models CO60 (upper) and CO100 (lower). The line features
seen in the synthetic spectra for CO60 are much too narrow compared with
observations, while the fits are much improved with CO100.
}
\label{fig:sp97ef}
\end{figure}

The light curve tail starts only $\sim 40$ days after maximum, much
later than in other SNe Ic. This can be reproduced by different
explosion models.  In Figure~\ref{fig:lc97ef} we compare the calculated V
light curves for the standard energy model CO60 and the energetic
model CO100 ($E = 10^{52}$erg) with the observed V light curve of
SN1997ef.  We adopt a distance of 52.3 Mpc (a distance modulus of
$\mu=33.6$ mag) as estimated from the recession velocity, 3,400 km
s$^{-1}$ (Garnavich et al. 1997) and a Hubble constant $H_0=65$ km
s$^{-1}$ Mpc$^{-1}$.  We assume no color excess, $E(B-V)=0.00$.  The
light curve of SN 1997ef has a very broad maximum, which lasts for
$\sim$ 25 days.

Since the model parameters of CO100 and CO60 give similar timescales,
the light curves of the two models look similar: both have quite a broad
peak and reproduce the light curve of SN1997ef 
reasonably well (Fig.~\ref{fig:lc97ef}).

The light curve shape depends also on the distribution of $^{56}$Ni,
which is produced in the deepest layers of the ejecta. More extensive
mixing of $^{56}$Ni leads to an earlier rise of the light curve.
For SN 1997ef, the best fit is obtained when $^{56}$Ni is mixed almost
uniformly to the surface for both models.  Without such extensive
mixing, the rise time to V $=$ 16.5 mag would be $\sim$ 30 d for CO100.
However, spectroscopic dating suggests that the peak occurred $\sim
18$ days after the explosion.

Model CO60 has the same kinetic energy ($E_{\rm K} = 1 \times 10^{51}$
erg) as model CO21, which was used for SN Ic 1994I (see Table 1 for the
model parameters). Since the light curve of SN 1997ef is much slower
than that of SN 1994I, the ejecta mass of CO60 is $\sim$ 5 times larger
than that of CO21.

The ejecta mass of CO100 is a factor of $\sim 2$ larger than that of
CO60, and it is only $\sim 20$\% smaller than that of model CO138,
which was used for SN 1998bw (Table 1).  Thus the explosion energy of
CO100 should be $\sim 10$ times larger than that of CO60 to reproduce
the light curve of SN 1997ef. This explosion is very energetic, but
still much weaker than the one in CO138.  The smaller $E_{\rm K}$ for
a comparable mass allows CO100 to reproduce the light curve of SN
1997ef, which has a much broader peak than that of SN 1998bw.

The light curve of SN 1997ef enters the tail around day 40.  Since
then, the observed V magnitude declines linearly with time at a rate
of $\sim 1.1 \times 10^{-2}$ mag day$^{-1}$, which is slower than in
other SNe Ic and is close to the $^{56}$Co decay rate $9.6 \times
10^{-3}$ mag day$^{-1}$.  Such a slow decline implies much more
efficient $\gamma$-ray trapping in the ejecta of SN 1997ef than in SN
1994I.  The ejecta of both CO100 and CO60 are fairly massive and are
able to trap a large fraction of the $\gamma$-rays, so that the
calculated light curves have slower tails compared with CO21.

However, the light curves of both models decline somewhat faster in
the tail than the observations.  A similar discrepancy has been noted
for the Type Ib supernovae (SNe Ib) 1984L and 1985F (Swartz \& Wheeler
1991; Baron, Young, \& Branch 1993).  The late time light curve
decline of these SNe Ib is as slow as the \co\ decay rate, so that the
inferred value of $M$ is significantly larger (and/or $E_{\rm K}$ is
smaller) than those obtained by fitting the early light curve shape.
Baron et al. (1993) suggested that the ejecta of these SNe Ib must be
highly energetic and as massive as $\sim$ 50 \ms.  We will suggest
that such a discrepancy between the early- and late-time light curves
might be an indication of asphericity in the ejecta of SN 1997ef and
that it might be the case in those SNe Ib as well.

\subsection{Spectra}

As we have shown, light curve modeling provides direct constraints on
$M_{\rm CO}$ and $E$.  However, it is difficult to determine uniquely
these values, and hence the characteristics of the explosion, from the
light curve shape alone, since models with different values of $M_{\rm
ej}$ and $E$ can yield similar light curves.  Fortunately, however,
models with different values of $M_{\rm ej}$ and $E$ are expected to
show different spectral evolution.

Using detailed spectrum synthesis, we can therefore distinguish
between different models clearly, because the spectrum
contains much more information than a single-band light curve.

Around maximum light, the spectra of SN~1997ef show just a few very
broad features, and are quite different from those of ordinary SNe Ib/c,
but similar to those of SN~1998bw.  However, at later epochs the spectra
develop features that are easy to identify, such as the Ca~II IR
triplet at $\sim 8200$\AA, the O~I absorption at 7500 \AA, several
Fe~II features in the blue, and they look very similar to the spectrum
of the ordinary SN Ic 1994I.

We computed synthetic spectra with a Monte Carlo spectrum synthesis
code using the density structure and composition of the hydrodynamic
models CO60 and CO100.
We produced synthetic spectra for three epochs near maximum, of
SN~1997ef: Nov 29, Dec 5, and Dec 17. These are early enough that the
spectra are very sensitive to changes in the kinetic energy.  As in
the light curve comparison, we adopted a distance modulus of
$\mu=33.6$ mag, and $E(B-V)=0.0$.

In Figure \ref{fig:sp97ef} (upper) we show the synthetic spectra
computed with the ordinary SN~Ic model CO60.  The lines in the spectra
computed with this model are always much narrower than the observations.
This clearly indicates a lack of material at high velocity in model CO60,
and suggests that the kinetic energy of this model is much too small.

Synthetic spectra obtained with the hypernova model CO100 for the same 3
epochs are shown in Figure \ref{fig:sp97ef} (lower).  The spectra show
much broader lines, and are in good agreement with the observations.
In particular, the blending of the Fe lines in the blue, giving rise to
broad absorption troughs, is well reproduced.  The two `emission peaks'
observed at $\sim 4400$ and 5200\AA\ correspond to the only two
regions in the blue that are relatively line-free.

The spectra are characterized by a low temperature, even near maximum,
because the rapid expansion combined with the relatively low
luminosity (from the tail of the light curve we deduce that SN~1997ef
produced about $0.15 M_\odot$ of $^{56}$Ni, compared to about $0.6
M_\odot$ in a typical SN~Ia and $0.5 M_\odot$ in SN~1998bw) leads to
rapid cooling.  Thus the \SiII\ 6355\AA\ line is not very strong.

Model CO100 has $E = 10^{52}$ erg, $M_{\rm ej} = 7.5 M_\odot$,
$M(^{56}$Ni) $= 0.15 M_\odot$.  From these values, we find $M_{\rm CO} =
10 M_\odot$, $M_{\rm rem} = 2.5 M_\odot$.  A $10 M_\odot$ CO core is
formed in a $\sim 30 M_\odot$ star.  Although model CO100 yields rather
good synthetic spectra, it still fails to reproduce the observed large
width of the O~I - Ca~II feature in the only near-maximum spectrum that 
extends sufficiently far to the red (5 Dec 1997). An improvement can be
obtained by introducing an arbitrary flattening of the density profile
at the highest velocities (Mazzali et al. 2000; Branch 2001). This
leads to higher values of both $E$ and $M_{\rm ej}$.

\subsection{Possible Aspherical Effects}

The light curve, the photospheric velocities, and the spectra of
SN~1997ef are better reproduced with the hyper-energetic model CO100
than with the ordinary SN Ic model CO60. However, there remain several
features that are still difficult to explain with model CO100.

These discrepancies may be interpreted as a possible sign of
asphericity in the ejecta: A part of the ejecta moves faster than
average to form the lines at high-velocities at early phases, while
the other part of ejecta expands with a lower velocity so that the
low-velocity Si II line comes up at later epochs. Having a
low-velocity component would also make it easier to reproduce the slow
tail and the long duration of the photospheric phase.

Extensive mixing of $^{56}$Ni is required to reproduce the short rise
time of the light curve.  One possibility to induce such mixing in the
velocity space is an asymmetric explosion.  In the extremely
asymmetric cases, material ejection may take place in a jet-like form
(e.g., MacFadyen \& Woosley 1999; Khokhlov et al. 1999).  A jet could
easily bring some Ni from the deepest layers out to the high velocity
surface.  The lack of a strong case of coincidence with a GRB suggests
that if a jet was produiced it was either weak or it was not pointing
towards us.

Unlike other hypernovae, SN~1997ef does not seem to be a unique case.
At least two other SNe, SN~1997dq (Matheson et al. 2001) and SN~1999ey
show very similar properties. Unfortunately, these two objects were
not very intensively observed. Whether this is just a coincidence, or
whether it really indicates that SN~1997ef-like hypernovae are more
frequent than others remains an interesting question that must be
answered observationally.

\section{SN~2002ap}

\begin{figure}[t]
 \begin{center}
  \includegraphics[width=.7\textwidth]{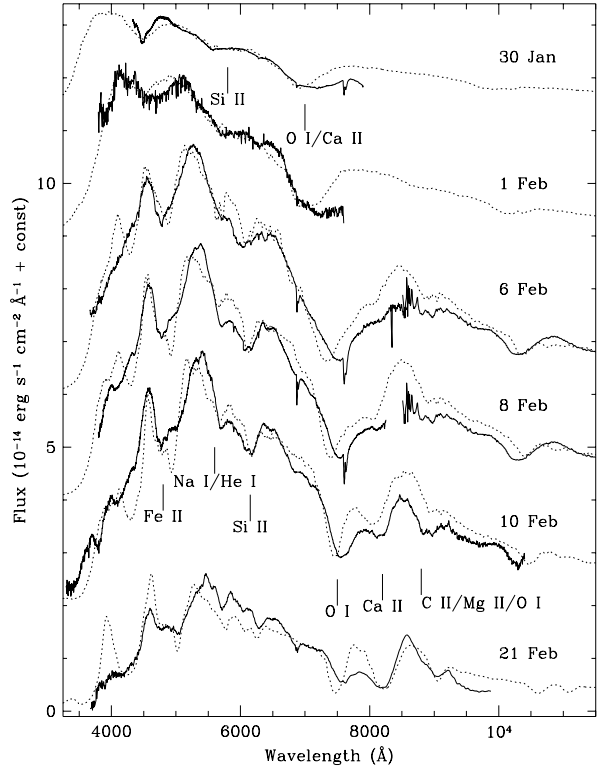}
 \end{center}
 \caption[] {A comparison between some observed spectra of SN 2002ap
({\em thick lines}: January 30 --- WHT; February 1 --- Gunma Obs.;
February 6 --- Beijing Obs.; February 8 --- Subaru FOCAS; February 7 IR ---
Subaru CISCO, shown twice; February 10 --- Lick Obs.; February 21 ---
Asiago Obs.) and synthetic spectra computed with model CO100/4 ({\em
dashed lines}) (Mazzali et al. 2002).
}
\label{sp02ap}
\end{figure}

SN~Ic 2002ap was discovered in M74 on 2002 January 30 (Hirose 2002).
The SN was immediately recognised as a hypernova from its broad
spectral features (Kinugasa et al. 2002; Meikle et al. 2002; Gal-Yam, 
Ofek \& Shemmer 2002; Filippenko \& Chornock 2002).  This indicates high
velocities in the ejected material, which is the typical signature of
hypernovae.  It was therefore followed from several observatories, and
the relative proximity also favored observations with small
telescopes.  Luckily, the SN was discovered very soon after it
exploded: the discovery date was January 29, while the SN was not
detected on January 25 (Nakano, Kushida, \& Li 2002).  This is among
the earliest any SN has been observed, with the obvious exceptions of
SN~1987A and SN~1993J.

\subsection{Light Curve}

Figure \ref{eps2} shows the $V$-band light curves of the same four SNe
as in Figure \ref{eps1}.  SN~2002ap reached $V$ maximum on about
February 8 at $V = 12.3$ mag.  SN~2002ap peaks earlier than both
hypernovae 1998bw and 1997ef, but later than the normal SN~1994I,
suggesting an intermediate value of the ejecta mass \Mej.

Using a distance to M74 of 8~Mpc ($\mu = 29.5$ mag; Sharina,
Karachentsev, \& Tikhonov 1996), and a combined Galaxy and M74
reddening of $E(B-V) = 0.09$ mag (estimated from a Subaru HDS
spectrum; Takada-Hidai, Aoki, \& Zhao 2002), 
the absolute magnitude is $M_V =
-17.4$.  This is comparable to SN~1997ef and fainter than SN~1998bw by
almost 2 mag.  Since peak brightness depends on the ejected \Nifs\
mass, SNe~2002ap, 1997ef, and 1994I appear to have synthesized similar
amounts of it.  Estimates were $\sim 0.07$ \Msun\ for SN~1994I (Nomoto
et al. 1994) and 0.13 \Msun\ for SN~1997ef (Mazzali et al. 2000).  The
$^{56}$Ni mass for SN~2002ap is estimated to be $\sim 0.07$ \Msun,
which is similar to that of normal core-collapse SNe such as SNe 1987A
and 1994I.

\subsection{Spectra}

Figure \ref{eps1} shows the maximum-light spectra of SN~2002ap, of the
hypernovae SNe 1998bw and 1997ef, and of the normal SN~Ic 1994I.  If
line width is the distinguishing feature of a hypernova, then clearly
SN~2002ap is a hypernova, as its spectrum resembles that of SN~1997ef
much more than that of SN~1994I.  Line blending in SN~2002ap and
SN~1997ef is comparable.  However, some individual features that are
clearly visible in SN~1994I but completely blended in SN~1997ef can at
least be discerned in SN~2002ap (\eg the \NaI--\SiII\ blend near
6000~\AA\ and the \FeII\ lines near 5000~\AA).  Therefore,
spectroscopically SN~2002ap appears to be located just below SN~1997ef
in a ``velocity scale," but far above SN~1994I, which appears to
confirm the evidence from the light curve.

The spectral evolution of SN~2002ap shown in Figure~\ref{sp02ap} 
appears to follow closely that of
SN~1997ef, at a rate about 1.5 times faster.  The spectra and the
light curve of SN 2002ap can be well reproduced by a model with
ejected heavy-element mass \Mej = 2.5--5 \Msun\ and $E_{51}=4$--10.
Both \Mej\ and \KE\ are much smaller than those of SNe 1998bw and
1997ef (but they could be larger if a significant amount of He is
present).

\subsection{Is SN~2002ap a Hypernova or a Supernova? }

Although SN~2002ap appears to lie between normal core-collapse SNe and
hypernovae, it should be regarded as a hypernova because its kinetic
energy is distinctly higher than for normal core-collapse SNe.  In
other words, the broad spectral features that characterize hypernovae
are the results of a high kinetic energy.  Also, SN~2002ap was not
more luminous than normal core-collapse SNe.  Therefore brightness
alone should not be used to discriminate hypernovae from normal SNe,
while the criterion should be a high kinetic energy, accompanied by
broad spectral features. Further examples of hypernovae are necessary
in order to establish whether a firm boundary between the two groups
exists.

For these values of \KE, \Mej, and $M(^{56}$Ni), we can constrain the
progenitor's main-sequence mass $M_{\rm ms}$ and the remnant mass
$M_{\rm rem}$.  Modeling the explosions of C+O stars with various
masses, we obtain $M(^{56}$Ni) as a function of the parameter set
(\KE, $M_{\rm CO}$, $M_{\rm rem} = $\Mej$ - M_{\rm CO}$).  The model
which is most consistent with our estimates of (\Mej, $E$) is one with
$M_{\rm CO} \approx 5$\,\Msun, $M_{\rm rem} \approx 2.5$\,\Msun, and
$E_{51} = 4.2$.  The 5 $M_\odot$ C+O core forms in a He core of mass
$M_\alpha = 7$ \Msun, corresponding to a main-sequence mass $M_{\rm
ms} \approx 20$--25 \Msun.  The $M_{\rm ms} - M_\alpha$ relation
depends on convection and metallicity (e.g., Nomoto \& Hashimoto 1988;
Umeda \& Nomoto 2002).

The estimated progenitor mass and explosion energy are both smaller
than those of previous Hypernovae such as SNe 1998bw and 1997ef, but
larger than those of normal core-collapse SNe such as SN 1999em.  This
mass range is consistent with the non-detection of the progenitor in
pre-discovery images of M74 (Smartt et al. 2002).

Given the estimated mass of the progenitor, binary interaction
including the spiral-in of a companion star (Nomoto et al. 2001a) is
probably required in order for it to lose its hydrogen and some (or
most) of its helium envelope.  This would suggest that the progenitor
was in a state of high rotation.  It is possible that a high rotation
rate and/or envelope ejection are also necessary conditions for the
birth of a hypernova.

\subsection{Possible Aspherical Effects}

SN~2002ap was not apparently associated with a GRB.  This may actually
be not so surprising, since the explosion energy of SN~2002ap is about
a factor of 5-10 smaller than that of SN~1998bw, as also indicated by
the weak radio signature (Berger, Kulkarni, \& Chevalier 2002).  The
present data show no clear signature of asymmetry, except perhaps for
some polarization (Kawabata et al. 2002; Leonard et al. 2002; Wang et
al. 2003), which is smaller than that of SN~1998bw.  This suggests
that the degree of asphericity is smaller in SN~2002ap and that the
possible ``jet" may have been weaker, which makes GRB generation more
difficult.

\section{SN~1999as}

SN~1999as was discovered on February 18, 1999 by the Supernova
Cosmology Project (Knop et al. 1999) in an anonymous galaxy having a
redshift of 0.127.  The absolute magnitude was exceptionally bright,
$M_{\rm V} < - 21.5$, at least nine times brighter than the hypernova
SN~1998bw.

SN~1999as is spectroscopically classified as a SN Ic because its
photospheric phase spectra show no conspicuous lines of hydrogen,
He~I, or Si II $\lambda$6355.  The usual SN~Ic spectral lines such as
Ca~II and O~I are very broad, like in other hypernovae.  However, some
narrow ($\sim$~2000 km~s$^{-1}$) but highly blueshifted ($\sim$ 11,000
km~s$^{-1}$) lines of Fe~II are also present (Hatano et al. 2001).

By fitting the observed light curve, we have obtained the following
constraints on the explosion model: the ejected mass $M_{\rm ej}
\simeq$ 10 -- 20 $M_{\odot}$, the kinetic energy of ejected matter $E
\simeq$ 10$^{52}$ -- 10$^{53}$ ergs, and the mass of ejected
radioactive $^{56}$Ni $M_{\rm Ni}$ $>$ 4 $M_{\odot}$ (Deng et
al. 2001). The progenitor of this may have been as massive as $\sim 60
M_{\odot}$, and the explosion almost certainly resulted in the
formation of a Black Hole.  Unfortunately, the spectral coverage is
not very extensive, and an accurate determination of the properties of
this supernova is therefore difficult.

The asymmetric hydrodynamical model of Maeda et al. (2002) could
represent this new class of hypernovae. In this model, a clump of
freshly synthesized $^{56}$Ni exists at high velocity ($\sim$~15,000
km~s$^{-1}$) near the symmetry axis; such a clump could produce narrow
but high--velocity absorption lines if the viewing angle with respect
to the symmetry axis is small. On the other hand, the lack of a
detected GRB may suggest that the angle was not very small.

\section{Type IIn Hypernova: SN~1997cy and SN~1999E}

SN~1997cy is a different type of hypernova, as it is of Type IIn.
However, the energy deduced from its light curve is extremely large.
The energy reveals itself in the strength of the ejecta-CSM
interaction. Furthermore, the SN may have a correlated GRB.  SN~1999E
is very similar to SN~1997cy in the spectra and light curve (Rigon et
al. 2003).

\begin{figure}[t]
 \begin{center}
  \includegraphics[width=.5\textwidth]{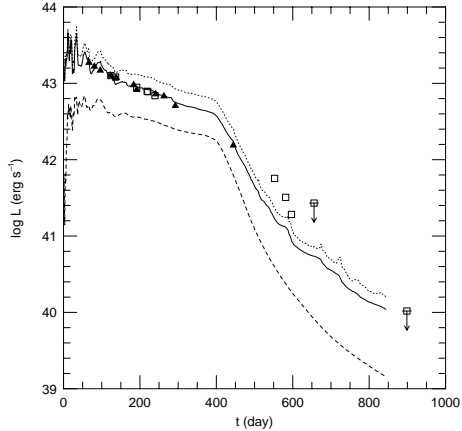}
 \end{center}
 \caption[] {The {\sl uvoir}  bolometric light curve of SN~1997cy
compared with the synthetic light curve obtained with the CSM
interaction model (Turatto et al. 2000). }
\label{fig:97cylc}
\end{figure}

SN 1997cy displayed narrow H$\alpha$ emission on top of broad wings,
which lead to its classification as a Type IIn (Germany et al. 2000;
Turatto et al. 2000).  Assuming $A_V=0.00$ for the galactic extinction
(NED) we get an absolute magnitude at maximum $M_v \le-20.1$.  It is
the brightest SN II discovered so far.  The light curve of SN 1997cy
does not conform to the classical templates of SN II, namely Plateau
and Linear, but resembles the slow evolution of Type IIn SN 1988Z.  As
seen from the {\it uvoir} bolometric light curve in Figure
\ref{fig:97cylc}, the SN light curve decline is slower than the
$^{56}$Co decay rate between day 120 to 250, suggesting circumstellar
interaction for the energy source.  (Here the outburst is taken to be
coincident with GRB970514.)

In the interaction model (Turatto et al. 2000), collision of the SN
ejecta with the slowly moving circumstellar matter (CSM) converts the
kinetic energy of the ejecta into light, thus producing the observed
intense light display of the SN.  The exploratory model considers the
explosion of a massive star of $M = 25 M_\odot$ with a parameterized
kinetic energy $E$.  The collision is assumed to start near the
stellar radius at a distance $r_1$, where the density of the CSM is
$\rho_1$, and adopt for the CSM a power-law density profile $\rho
\propto r^{n}$.  The parameters $E_{\rm K}$, $\rho_1$, and $n$, are
constrained from comparison with the observations.

The regions excited by the forward and reverse shock emit mostly
X-rays. The density in the shocked ejecta is so high that the reverse
shock is radiative and a dense cooling shell is formed (e.g., Suzuki
\& Nomoto 1995; Terlevich et al. 1992).  The X-rays are absorbed by
the outer layers and the core of the ejecta, and re-emitted as
UV-optical photons.

Narrow lines are emitted from the slowly expanding unshocked CSM
photoionized by the SN UV outburst or by the radiation from the
shocks; intermediate width lines come from the shock-heated CSM; broad
lines come from either the cooler region at the interface between
ejecta and CSM.  

Figure \ref{fig:97cylc} shows the model light curve which best fits
the observations.  The model parameters are: $E = 3\times10^{52}$ erg,
$\rho_1 = 4\times10^{-14}$ g cm$^{-3}$ at $r_1 = 2 \times10^{14}$ cm
(which corresponds to a mass-loss rate of $\dot{M}=4\times10^{-4}$
$M_{\odot}$ yr$^{-1}$ for a wind velocity of 10 \kms), and $n = -1.6$.
The large mass-loss episode giving rise to the dense CSM is supposed
to occur after the progenitor makes a loop in the HR diagram from BSG
to RSG.  In this model, the mass of the low-velocity CSM is $\sim 5
M_\odot$, which implies that the transition from BSG to RSG took place
about $10^4$ yr before the SN event.

The large CSM mass and density are necessary to have large shocked
masses and thus to reproduce the observed high luminosity, and so is
the very large explosion energy.  For models with low $E_{\rm K}$ and
high $\rho_1$, the reverse shock speed is too low to produce a
sufficiently high luminosity.  For example, a model with $E = 10^{52}$
erg and $\rho_1$ as above yields a value of $L_{\rm UVOIR}$ lower than
the observed luminosity by a factor of $\sim$ 5. For high $E$ or low
$\rho_1$, the expansion of the SN ejecta is too fast for the cooling
shell to absorb enough X-rays to sustain the luminosity.  Thus in this
model $E_{\rm K}$ and $\dot{M}$ are constrained within a factor of
$\sim$ 3 of the reported values.

The shape of the light curve constrains the circumstellar density
structure. For $n = -2$, the case of a steady wind, $L_{\rm UVOIR}$
decreases too rapidly around day 200. To reproduce the observed
decrease after day $\sim$ 300, the CSM density is assumed to drop
sharply at the radius the forward shock reaches at day 300, so that
the collision becomes weaker afterwards.  (Such a change of the CSM
density corresponds to the transition from BSG to RSG of the
progenitor $\sim 10^4$ yr before the SN explosion.)  This is
consistent with the simultaneous decrease in the H$\alpha$ luminosity.

The observed light curve drops sharply after day 550.  The model
reproduces such a light curve behavior (Figure \ref{fig:97cylc})
assuming that when the reverse shock propagates through $\sim$ 5
$M_\odot$, it encounters exceedingly low density region and thus it
dies.  In other words, the model for the progenitor of SN 1997cy
assumes that most of the core material has fallen into a massive black
hole of, say, $\sim 10 M_\odot$, while the extended H/He envelope of
$\sim$ 5 $M_\odot$ has not collapsed.  Then material is ejected from
the massive black hole possibly in a jet-like form, and the envelope
is hit by the ``jet'' and ejected at high velocity.

In this model, the ejecta are basically the H/He layers and thus
contain the original (solar abundance) heavy elements plus some heavy
elements mixed from the core (before fall back) or jet materials.
This might explain the lack of oxygen and magnesium lines in the
spectra particularly at nebular phases (Turatto et al. 2000).

\section{Properties of Hypernovae}\label{sec:discussion}

Based on the observed objects and their interpretation, it is possible
to make some generalisation regarding the properties of Hypernovae and
their relation to the progenitor stars.

\subsection{The Explosion Kinetic Energy}

In Figure \ref{fig:nimass} we plot $E$ as a function of the
main-sequence mass $M_{\rm ms}$ of the progenitor star as derived from
fitting the optical light curves and spectra of various hypernovae, of
the normal SNe 1987A, 1993J, and 1994I (e.g., Shigeyama \& Nomoto
1990; Nomoto et al. 1993; Nomoto et a. 1994; Shigeyama et al. 1994;
Iwamoto et al. 1994; Woosley et al. 1994; Young, Baron, \& Branch
1995), and of SNe 1997D (Turatto et al. 1998). 
Properties of Type Ib/c supernovae/hypernovae thus derived 
are summarized in Table~\ref{tab:hyperlist}. It appears
that $E$ increases with $M_{\rm ms}$, forming a `Hypernova Branch',
reaching values much larger than the canonical $10^{51}$ erg.
SNe~1997D and 1999br, on the contrary, are located well below that
branch, forming a 'Faint SN Branch'.

This trend might be interpreted as follows.  Stars with $M_{\rm ms}
\lsim$ 20-25 \ms\ form a neutron star (SN 1987A may be a borderline
case between the neutron star and black hole formation).  Stars with
$M_{\rm ms} \gsim$ 20-25 \ms\ form a black hole (e.g., Ergma \& van
den Heuvel 1998); whether they become hypernovae or faint SNe may
depend on the angular momentum in the collapsing core, which in turn
depends on the stellar winds, metallicity, magnetic fields, and
binarity.

Hypernovae might have rapidly rotating cores owing possibly to the
spiraling-in of a companion star in a binary system.  The core of faint
SNe II might not have a large angular momentum, because the progenitor
had a massive H-rich envelope so that the angular momentum of the core
might have been transported to the envelope possibly via a
magnetic-field effect.

Between these two branches, there may be a variety of SNe.  A
dispersion in the properties of SNe II-P has been reported (Hamuy
2003).

\subsection{The Mass of Ejected $^{56}$Ni }

\begin{figure}[t]
 \begin{center}
  \includegraphics[width=.6\textwidth]{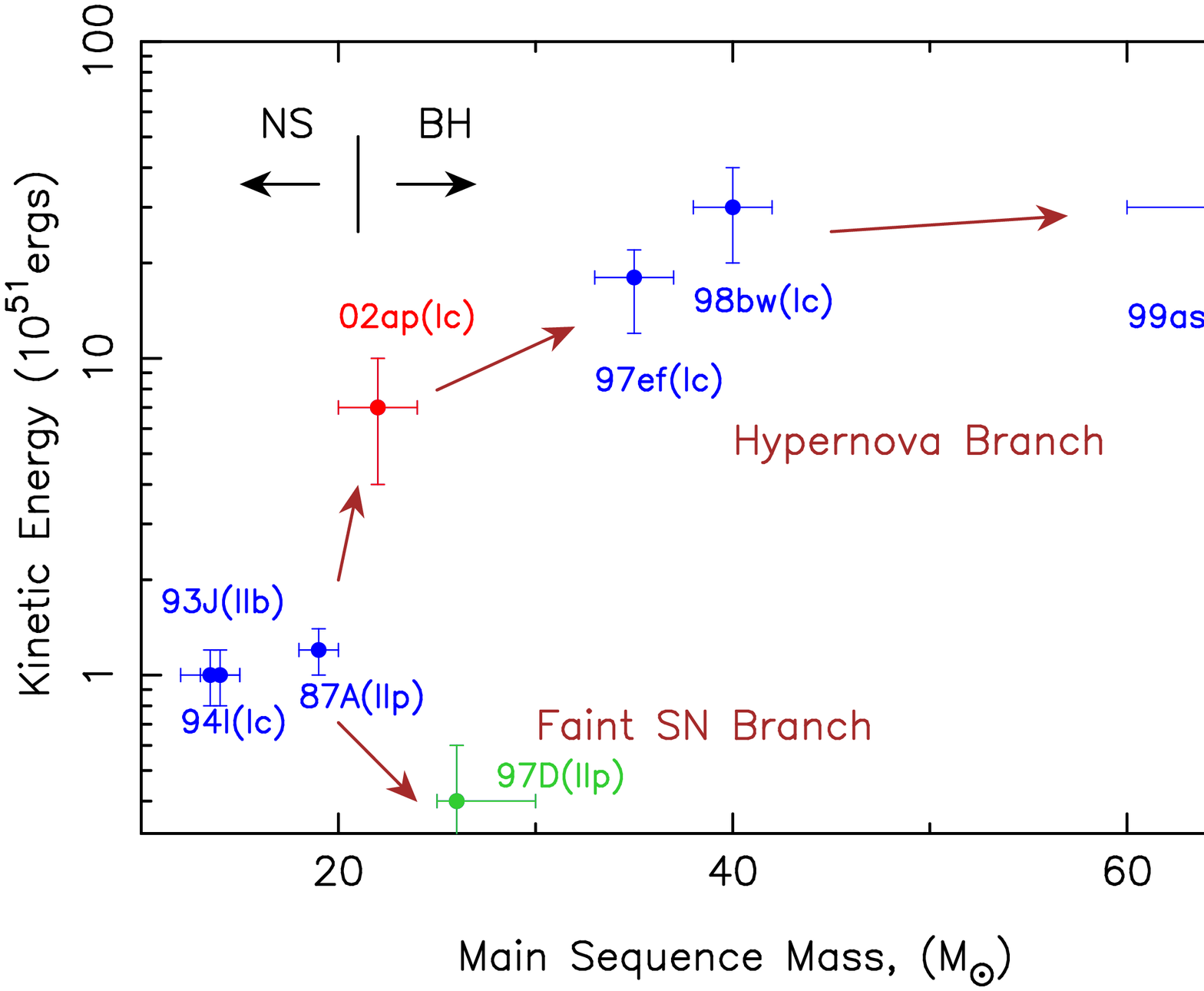}
\vskip 0.2cm
  \includegraphics[width=.6\textwidth]{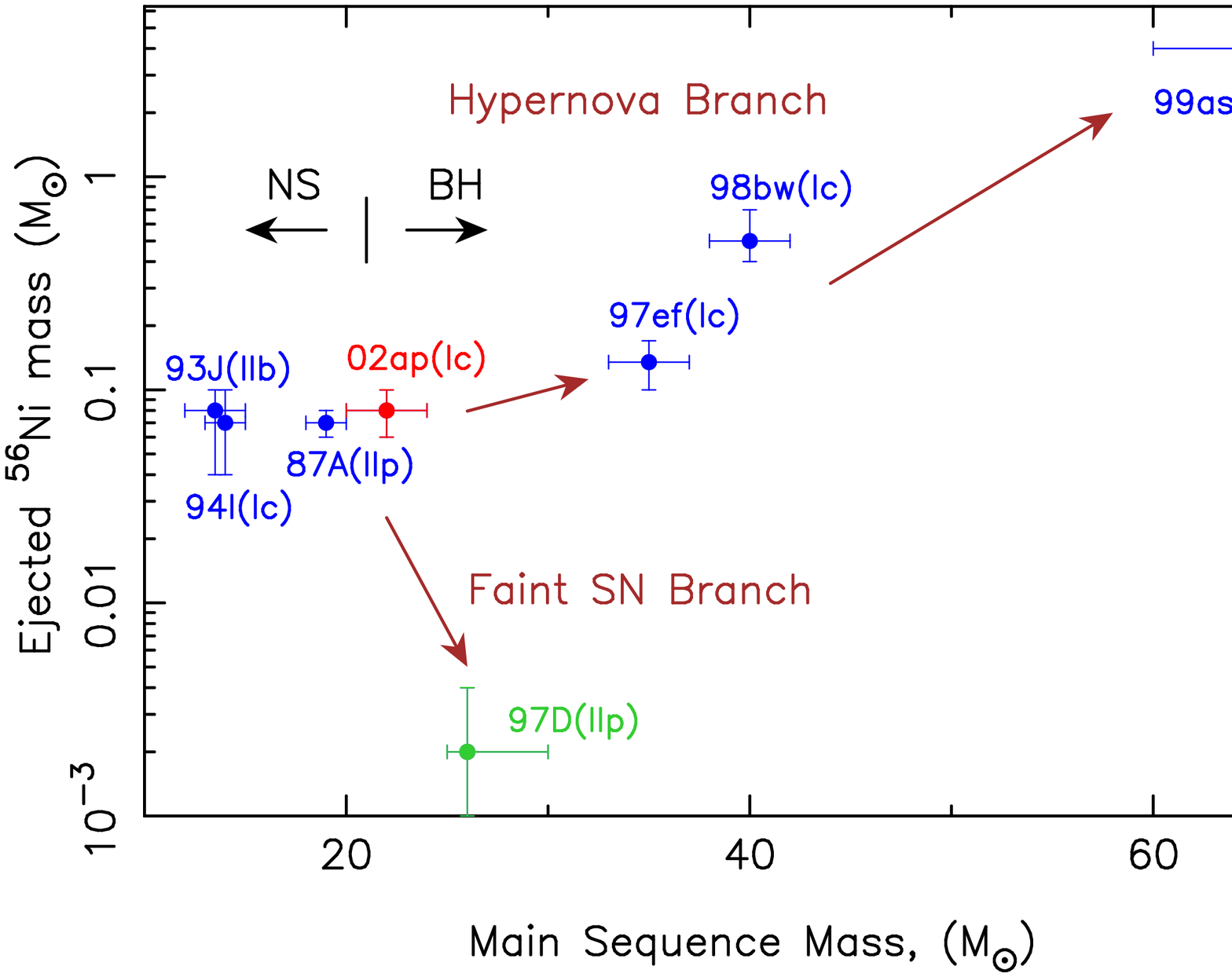}
 \end{center}
 \caption[] {Explosion energies and the ejected $^{56}$Ni mass
against the main sequence mass of the
progenitors for several bright supernovae/hypernovae (Nomoto et al. 2003).
\label{fig:nimass}}
\end{figure}

A similar relation is observed between the mass of $^{56}$Ni,
$M(^{56}$Ni), synthesized in core-collapse supernovae and $M_{\rm ms}$
in Figure \ref{fig:nimass}, which is important to know for the study
of the chemical evolution of galaxies.  Stars with $M_{\rm ms} \lsim$
20-25 \ms, forming a neutron star, produce $\sim$ 0.08 $\pm$ 0.03 \ms\
\Nifs\ as in SNe 1993J, 1994I, and 1987A.

For stars with $M_{\rm ms} \gsim$ 20-25 \ms, which form black holes,
$M(^{56}$Ni) appears to increase with $M_{\rm ms}$ in the `Hypernova
Branch', while SNe in the ' Faint SN Branch' produced only very little
$^{56}$Ni.  For faint SNe, because of the large gravitational potential,
the explosion energy was so small that most of \Nifs\ fell back onto a
compact star remnant.

\begin{table}[t]
\begin{center}
\begin{tabular}{cccccc} \hline
               & 83N & 94I & 02ap & 97ef & 98bw\\
               & Ib  & Ic  & \multicolumn{3}{c}{Ic (Hypernovae)}\\\hline
\multicolumn{6}{c}{Pre-explosion}\\
$M_{\rm ZAMS}$ & 15  & 15  & 21   & 34   & 40\\
$M_{\rm He}$   & 4   & 4   & 6.6  & 13   & 16\\
$M_{\rm CO}$   & 2   & 2   & 4.5  & 11   & 14\\
$M_{\rm exp}$  & 4   & 2.1 & 4.6  & 11.1 & 13.8\\\hline
\multicolumn{6}{c}{Post-explosion}\\
$M_{\rm rem}$  & 1.25& 1.2 & 2.1  & 1.6  & 2.9\\
$M_{\rm ej}$   & 2.75& 0.9 & 2.5  & 9.5  & 10.9\\
$M_{\rm He}$   & 2.0 & 0   & 0.1  & 0    & 0\\
$M_{\rm CO}$   & 0.5 & 0.6 & 1.8  & 5.3  & 8\\
$M_{\rm IME}$   & 0.1 & 0.2 & 0.5  & 4    & 2\\
$M_{\rm Ni}$   & 0.15& 0.07& 0.1  & 0.13 & 0.7\\
$E_{51}$       & 1   & 1   & 7    & 19   & 30\\\hline
\end{tabular}
\end{center}
\caption{Properties of supernovae and hypernovae.
\label{tab:hyperlist}}
\end{table}

\subsection{Asymmetry}

All Hypernovae of Type Ic show some signatures of asymmetry, or at
least of a departure from purely 1-dimensional spherically symmetric
models. This may support the case for their connection with at least
some GRB's.

So far only SN~1998bw and 2003dh have the well established connection
with GRBs.  In the other cases, either a GRB was not generated, or if
it was it was weak and/or not pointing towards us. The issue of
directionality is very important. If hypernovae are aspherical, we
expect to find a range of hypernova properties for the same Ni
mass. This can be established at late times, independetly of the shape
of the ejecta. These objects may be very different at early phases,
showing different light curves, velocities, abundances. So far,
however, this evidence is missing.

\subsection{Gamma-Ray Bursts/Supernovae Connection}

Candidates for the GRB/SN connection include GRB 980425/SN Ic 1998bw
(Galama et al. 1998; Iwamoto et al. 1998),
GRB 971115/SN Ic 1997ef (Wang \& Wheeler 1998),
GRB 970514/SN IIn 1997cy (Germany et al. 2000; Turatto et al. 2000),
GRB 980910/SN IIn 1999E (Thorsett \& Hogg1999), and
GRB 991002/SN IIn 1999eb (Terlevich, Fabian, \& Turatto 1999).
Recently, GRB 030329 has shown the evidence of a supernova (SN Ic 2003dh) 
in its optical afterglow spectra, confirming the GRB/SN connection directly 
(Stanek et al. 2003; Hjorth et al. 2003; Kawabata et al. 2003). 

Several GRB's are suggested to be associated with SNe, such as GRB 980326
(Bloom et al. 1999), GRB 970228 (Reichart 1999; Galama et al. 2000),
and GRB 011121/SN 2001ke (Bloom et al. 2002; Garnavich et al. 2002).
The decline of the light curve of the optical afterglows of these GRBs
slowed down at late phases, and this can be reproduced if a
red-shifted SN~1998bw-like light curve is superposed on the power-law
component.

A question is whether the supernovae associated with GRBs have a
uniform maximum luminosity, i.e., whether $\sim$ 0.5 \ms\ \Nifs\
production as in SN 1998bw is rather common or not.  Figure
\ref{fig:nimass} shows that the \Nifs\ mass and thus intrinsic maximum
brightness of Hypernovae has a large diversity.  We certainly need
more examples before we can define the luminosity function and the
actual distribution of masses of \(^{56}\)Ni produced in
supernovae/hypernovae.

For several hypernovae such as SNe 1998ey and 2002ap, no GRB counterpart
has been proposed.  These hypernovae were less energetic events than
SN~1998bw. It is possible that a weaker explosion is less efficient in
collimating the $\gamma$-rays to give rise to a detectable GRB
(GRB980425 was already quite weak compared to the average GRBs), or that
some degree of inclination of the beam axis to the line-of-sight results
in a seemingly weaker supernova and in the non-detection of a GRB. Only
the accumulation of more data will allow us to address these questions.

Properties of hypernova nucleosynthesis suggest that hypernovae of
massive stars may make important contributions to the Galactic (and
cosmic) chemical evolution (Nakamura et al. 2001b; Nomoto et
al. 2001b).  In view of small frequencies of GRBs, this implies that
hypernovae are much more frequent than GRBs, i.e., only a special
class of hypernovae gives rise to GRBs (Nomoto et al. 2003).

\section{Possible Evolutionary Scenarios to Hypernovae}

Here we classify possible evolutionary paths leading to C+O star
progenitors.  In particular, we explore the paths to the progenitors
that have rapidly rotating cores with a special emphasis, because the
explosion energy of hypernovae may be extracted from rapidly rotating
black holes (Blandford \& Znajek 1977).

(1) Case of a single star: If the star is as massive as $M_{\rm ms}
\gsim$ 40 \ms, it could lose its H and He envelopes in a strong stellar
wind (e.g., Schaller et al. 1992).  This would be a Wolf-Rayet star.

(2) Case of a close binary system: Suppose we have a close binary
system with a large mass ratio. In this case, the mass transfer from
star 1 to star 2 inevitably takes place in a non-conservative way, and
the system experiences a common envelope phase where star 2 is
spiraling into the envelope of star 1.  If the spiral-in releases
enough energy to remove the common envelope, we are left with a bare
He star (star 1) and a main-sequence star (star 2), with a reduced
separation.  If the orbital energy is too small to eject the common
envelope, the two stars merge to form a single star (e.g.,
van den Heuvel 1994).

(2-1) For the non-merging case, possible channels from the He stars to
the C+O stars are as follows (Nomoto, Iwamoto, \& Suzuki 1995).

(a) Small-mass He stars tend to have large radii, so that they can
fill their Roche lobes more easily and lose most of their He envelope
via Roche lobe overflow.

(b) On the other hand, larger-mass He stars have radii too small to
fill their Roche lobes.  However, such stars have large enough
luminosities to drive strong winds to remove most of the He layer
(e.g., Woosley, Langer, \& Weaver 1995).  Such a mass-losing He star
would again be a Wolf-Rayet star.

Thus, from the non-merging scenario, we expect two different kinds of
SNe Ic, fast and slow, depending on the mass of the progenitor.  SNe
Ic from smaller mass progenitors (channel 2-1-a) show faster light-curve
and spectral evolutions, because the ejecta become more quickly
transparent to both gamma-ray and optical photons. The slow SNe Ic
originate from the Wolf-Rayet progenitors (channels 1 and 2-1-b).  The
presence of both slow and fast SNe Ib/Ic has been noted by Clocchiatti
\& Wheeler (1997).

(2-2) For the merging case, the merged star has a large angular
momentum, so that its collapsing core must be rotating rapidly.  This
would lead to the formation of a rapidly rotating black hole from which
possibly a hyper-energetic jet could emerge.  If the merging process is
slow enough to eject the H/He envelope, the star would become a rapidly
rotating C+O star.  Such stars are the candidates for the progenitors of
Type Ic hypernovae like SNe 1997ef and 1998bw.  If a significant amount
of H-rich envelope remains after merging, the rapidly rotating core
would lead to a hypernova of Type IIn possibly like SN 1997cy (or Type Ib).

\section{Explosive Nucleosynthesis in Hypernovae}\label{sec:nuc}

\begin{figure}[t]
 \begin{center}
\begin{minipage}[t]{0.47\textwidth}
		\includegraphics[width=0.95\textwidth]{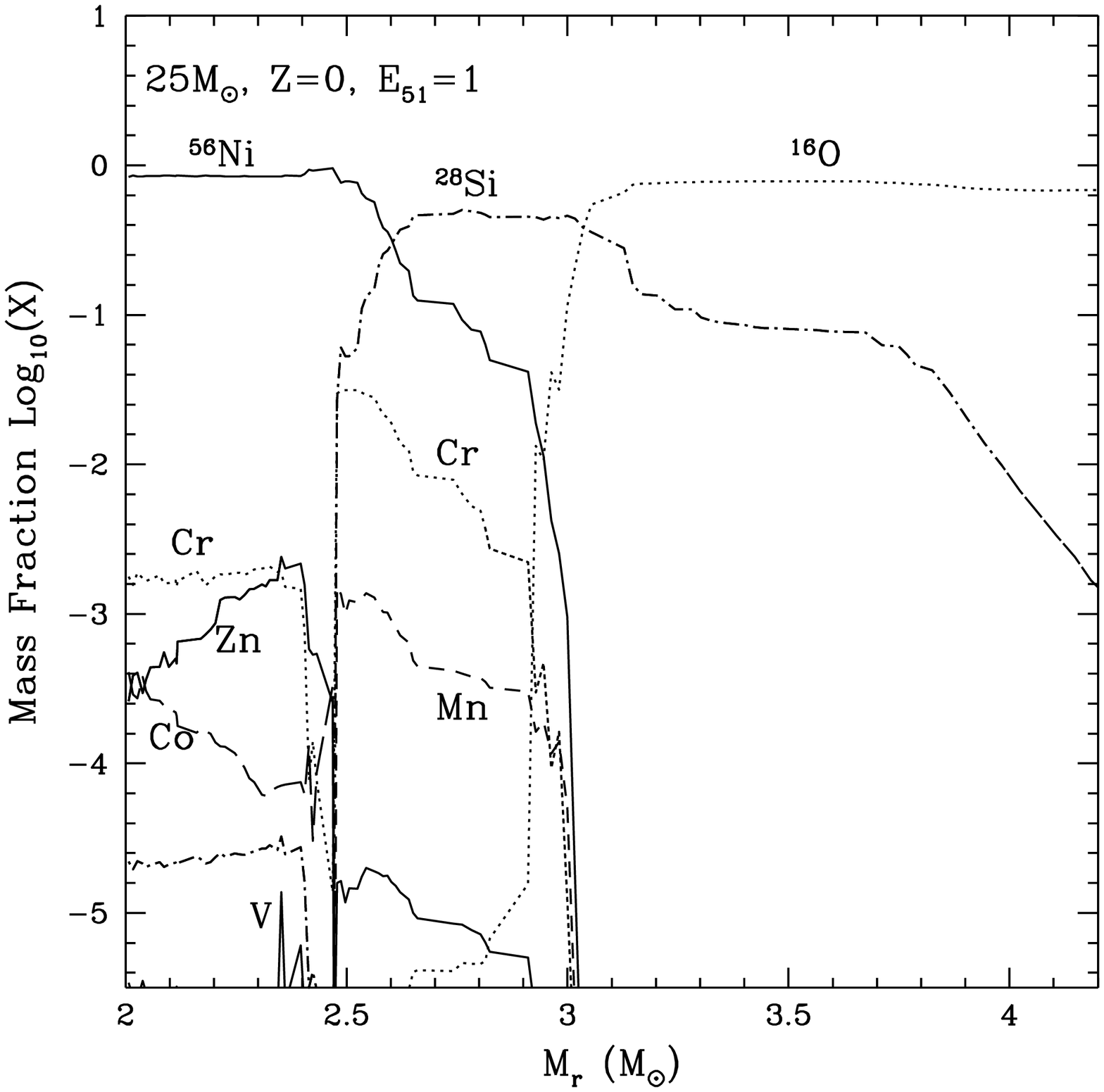}
\end{minipage}
\begin{minipage}[t]{0.47\textwidth}
		\includegraphics[width=0.95\textwidth]{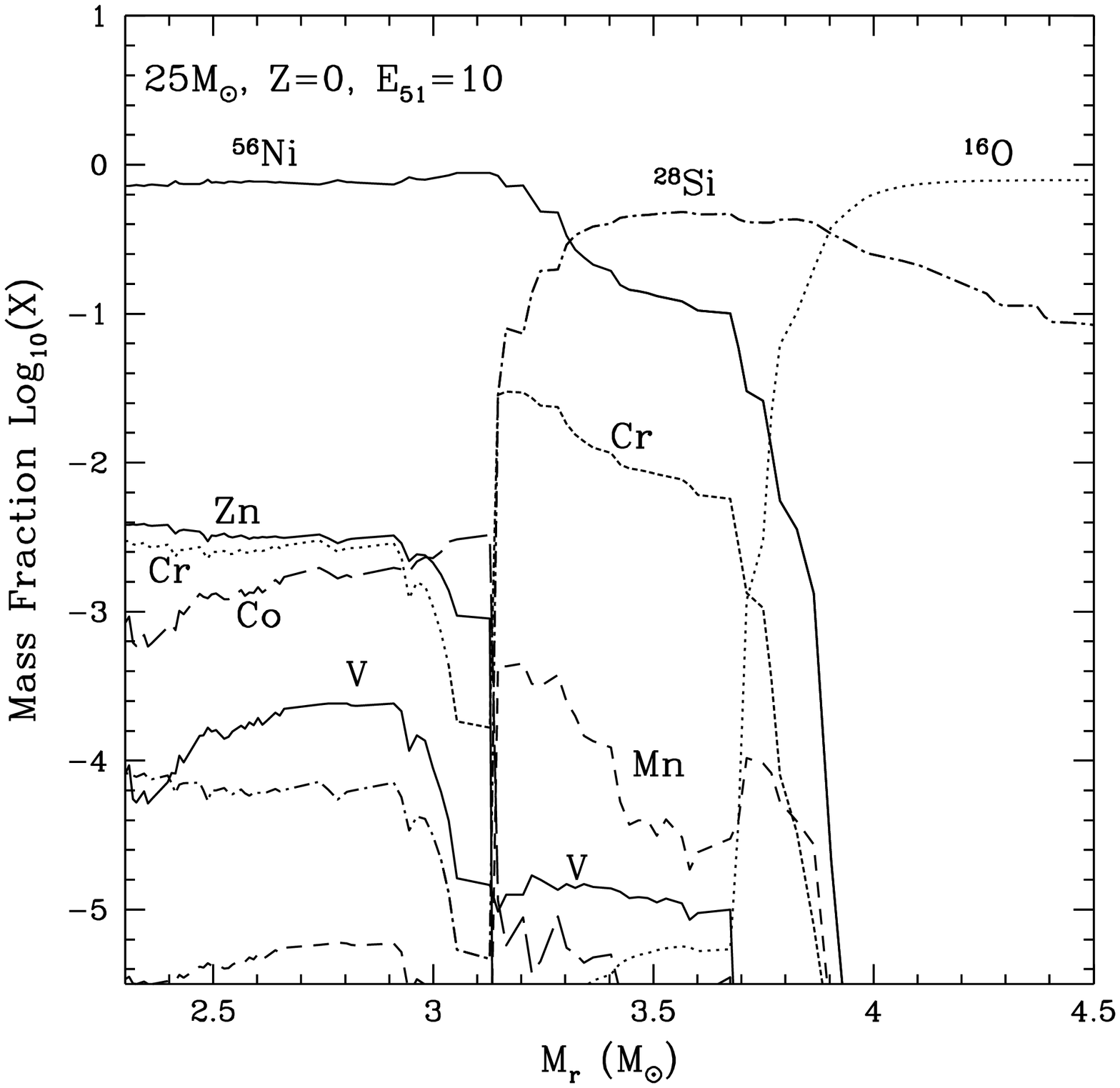}
\end{minipage}
 \end{center}
 \caption[]{Abundance distribution plotted against the enclosed mass
$M_r$ after the explosion of Pop III 25 \ms\ stars with $E_{51} = 1$
(left) and $E_{51} = 10$ (right) (Umeda \& Nomoto 2002). }
\label{fig:hnus}
\end{figure}

In core-collapse supernovae/hypernovae, stellar material undergoes
shock heating and subsequent explosive nucleosynthesis. Iron-peak
elements are produced in two distinct regions, which are characterized
by the peak temperature, $T_{\rm peak}$, of the shocked material.  For
$T_{\rm peak} > 5\times 10^9$K, material undergoes complete Si burning
whose products include Co, Zn, V, and some Cr after radioactive
decays.  For $4\times 10^9$K $<T_{\rm peak} < 5\times 10^9$K,
incomplete Si burning takes place and its after decay products include
Cr and Mn (e.g., Hashimoto, Nomoto, \& Shigeyama 1989; Thielemann,
Nomoto, \& Hashimoto 1996).

\subsection {Supernovae vs. hypernovae}

The right panel of Figure \ref{fig:hnus} shows the composition in the
ejecta of a 25 \ms\ hypernova model ($E_{51} = 10$).  The
nucleosynthesis in a normal 25 \ms\ SN model ($E_{51} = 1$) is also
shown for comparison in the left panel of Figure \ref{fig:hnus}.

We note the following characteristics of nucleosynthesis with very
large explosion energies (Nakamura et al. 2001b; Nomoto et al. 2001ab; 
Ohkubo, Umeda, \& Nomoto 2003):

(1) Both complete and incomplete Si-burning regions shift outward in
mass compared with normal supernovae, so that the mass ratio between
the complete and incomplete Si-burning regions becomes larger.  As a
result, higher energy explosions tend to produce larger [(Zn, Co,
V)/Fe] and smaller [(Mn, Cr)/Fe] 
([X/Y] $\equiv \log10 ({\rm X/Y}) - \log10({\rm X/Y})_{\odot}$), 
which can explain the trend observed
in very metal-poor stars (Ohkubo et al. 2003).

(2) In the complete Si-burning region of hypernovae, elements produced
by $\alpha$-rich freezeout are enhanced.  Hence, elements synthesized
through capturing of $\alpha$-particles, such as $^{44}$Ti, $^{48}$Cr,
and $^{64}$Ge (decaying into $^{44}$Ca, $^{48}$Ti, and $^{64}$Zn,
respectively) are more abundant.

(3) Oxygen burning takes place in more extended regions for the larger
KE.  Then more O, C, Al are burned to produce a larger amount of
burning products such as Si, S, and Ar.  Therefore, hypernova
nucleosynthesis is characterized by large abundance ratios of [Si,S/O],
which can explain the abundance feature of M82 
(Umeda et al. 2002).

\subsection{Hypernovae and Zn, Co, Mn, Cr}

\begin{figure}[t]
  \begin{center}
\begin{minipage}[t]{0.47\textwidth}
		\includegraphics[width=0.95\textwidth]{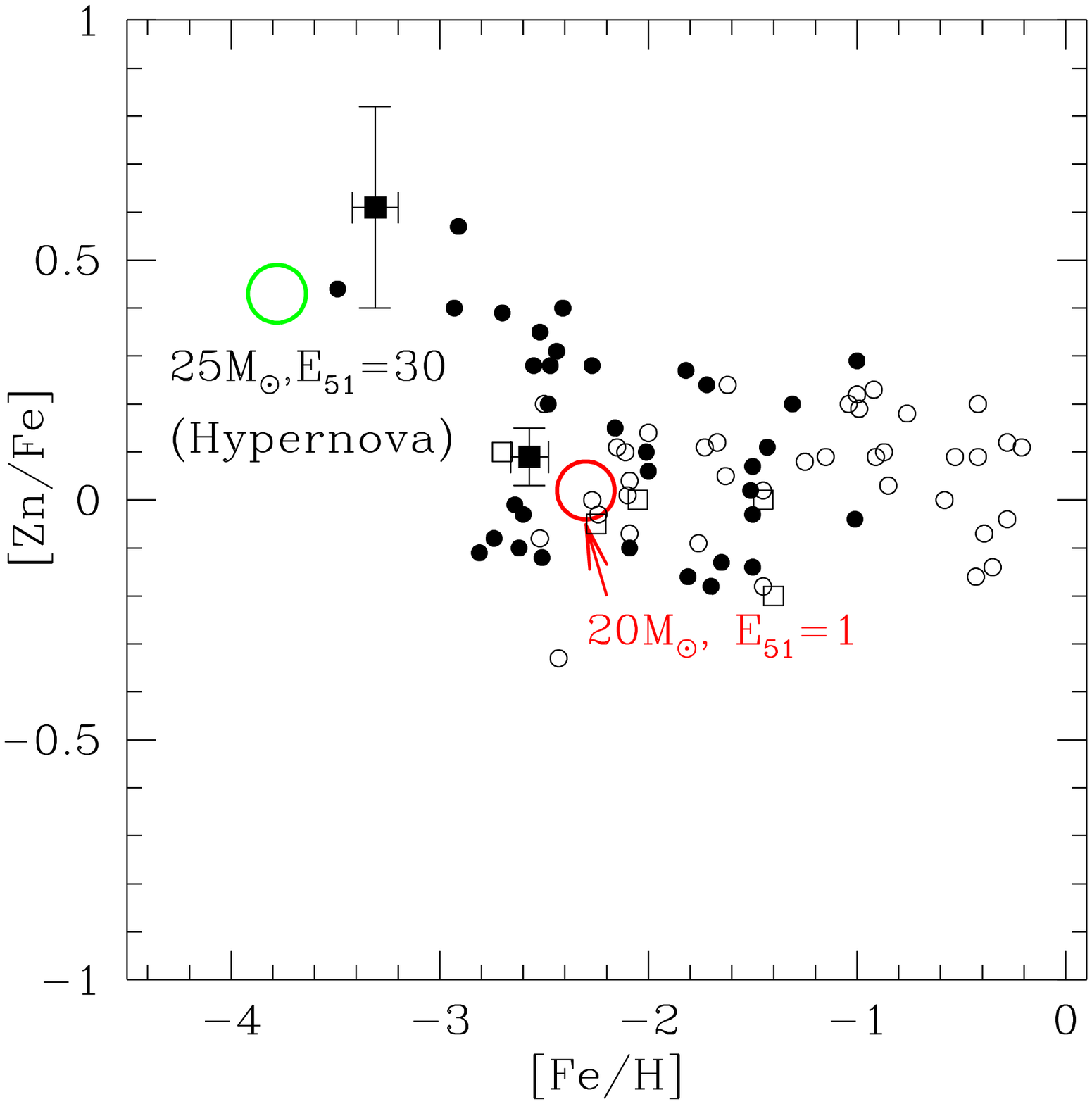}
  \end{minipage}
 \begin{minipage}[t]{0.47\textwidth}
		\includegraphics[width=0.95\textwidth]{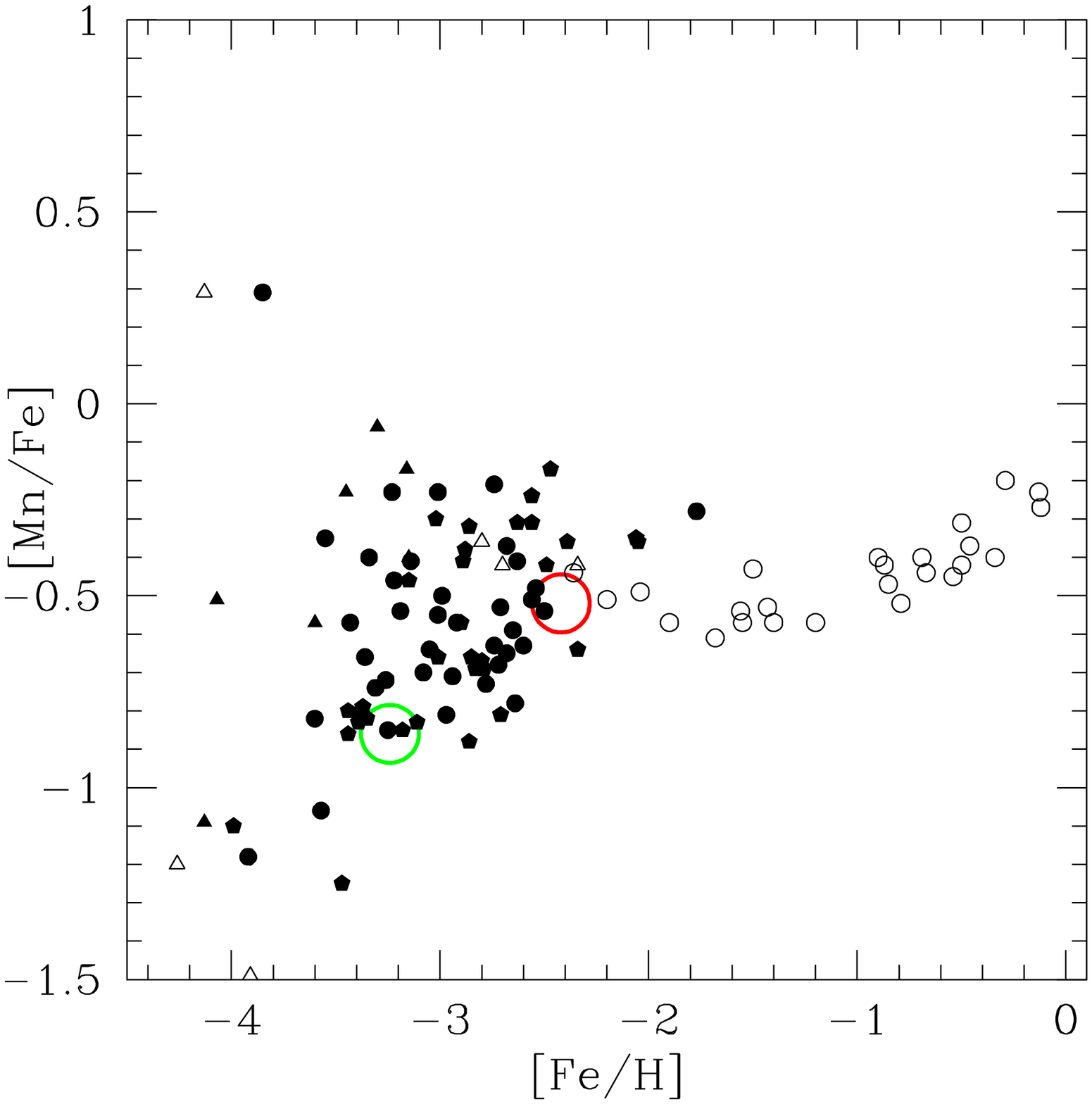}
  \end{minipage}  
\end{center}
\caption{
Observed abundance ratios of [Zn/Fe] and [Mn/Fe] and the theoretical
abundance patterns for a normal SN II (20$M_\odot$, $E_{51}=1$) and a
hypernova (25$M_\odot$, $E_{51}=30$) models (Ohkubo, Umeda, \& Nomoto 2003).
\label{fig:znfe}}
\end{figure}

Hypernova nucleosynthesis may have made an important contribution to
Galactic chemical evolution.  In the early galactic epoch when the
galaxy was not yet chemically well-mixed, [Fe/H] may well be
determined by mostly a single SN event (Audouze \& Silk 1995). The
formation of metal-poor stars is supposed to be driven by a supernova
shock, so that [Fe/H] is determined by the ejected Fe mass and the
amount of circumstellar hydrogen swept-up by the shock wave (Ryan,
Norris, \& Beers 1996).  Then, hypernovae with larger $E$ are likely
to induce the formation of stars with smaller [Fe/H], because the mass
of interstellar hydrogen swept up by a hypernova is roughly
proportional to $E$ (Ryan et al. 1996; Shigeyama \& Tsujimoto 1998)
and the ratio of the ejected iron mass to $E$ is smaller for
hypernovae than for normal supernovae.

In the observed abundances of halo stars, there are significant
differences between the abundance patterns in the iron-peak elements
below and above [Fe/H]$ \sim -2.5$ - $-3$.

(1) For [Fe/H]$\lsim -2.5$, the mean values of [Cr/Fe] and [Mn/Fe]
decrease toward smaller metallicity, while [Co/Fe] increases
(Fig. \ref{fig:znfe}: McWilliam et al. 1995; Ryan et al. 1996).

(2) [Zn/Fe]$ \sim 0$ for [Fe/H] $\simeq -3$ to $0$ (Sneden, Gratton,
\& Crocker 1991), while at [Fe/H] $< -3.3$, [Zn/Fe] increases toward
smaller metallicity (Fig. \ref{fig:znfe}: Primas et al. 2000; Blake et
al. 2001).

The larger [(Zn, Co)/Fe] and smaller [(Mn, Cr)/Fe] in the supernova
ejecta can be realized if the mass ratio between the complete Si
burning region and the incomplete Si burning region is larger, or
equivalently if deep material from the complete Si-burning region is
ejected by mixing or aspherical effects.  This can be realized if (1)
the mass cut between the ejecta and the compact remnant is located at
smaller $M_r$ (Nakamura et al. 1999), (2) $E$ is larger to move the
outer edge of the complete Si burning region to larger $M_r$ (Nakamura
et al. 2001b), or (3) mixing and/or asphericity in the explosion is
larger (Umeda \& Nomoto 2002, 2003; Maeda \& Nomoto 2003).

Among these possibilities, a large explosion energy $E$ enhances
$\alpha$-rich freezeout, which results in an increase of the local
mass fractions of Zn and Co, while Cr and Mn are not enhanced
(Umeda \& Nomoto 2002; Ohkubo et al. 2003).  Models with $E_{51} = 1 $ do
not produce sufficiently large [Zn/Fe].  To be compatible with the
observations of [Zn/Fe] $\sim 0.5$, the explosion energy must be much
larger, i.e., $E_{51} \gsim 20$ for $M \gsim 20 M_\odot$, i.e.,
hypernova-like explosions of massive stars ($M \gsim 25 M_\odot$) with
$E_{51} > 10$ are responsible for the production of Zn.

In the hypernova models, the overproduction of Ni, as found in the
simple ``deep'' mass-cut model, can be avoided (Ohkubo et al. 2003).
Therefore, if hypernovae made significant contributions to the early
Galactic chemical evolution, it could explain the large Zn and Co
abundances and the small Mn and Cr abundances observed in very
metal-poor stars (Umeda \& Nomoto 2002; Ohkubo et al. 2003). 

\subsection{Mixing and Fall-back}

As noted above, large [Zn, Co/Fe] and small [Mn, Cr/Fe] can be
obtained simultaneously if $M_{\rm cut}$ is sufficiently small and $E$
is sufficiently large.  However, the ejected $^{56}$Ni mass is larger 
for smaller $M_{\rm cut}$, and $M(^{56}$Ni) required to get
[Zn/Fe]$\sim 0.5$ appears to be too large to be compatible with
observations [O/Fe]$\sim 0 - 0.5$ in extremely metal-poor stars.

Here we consider a possible process that realizes effectively smaller
mass-cuts without changing the $^{56}$Ni mass.  In SNe II, when the
rapidly expanding core hits the H and He envelopes, a reverse shock
forms and decelerates core expansion.  The deceleration induces
Rayleigh-Taylor instabilities at the composition interfaces of H/He,
He/C+O, and O/Si as has been found in SN 1987A (e.g., Ebisuzaki,
Shigeyama, \& Nomoto 1989; Arnett et al. 1989).  Therefore, mixing can
take place between the complete and incomplete Si burning regions
according to the recent two dimensional numerical simulations
(Kifonidis et al. 2000).  The reverse shock can further induce matter
fall-back onto the compact remnant (e.g., Chevalier 1989).

Based on these earlier findings, we propose that the following
``mixing fall-back'' process takes place in most SNe II.  

(1) Burned material is uniformly mixed between the ``initial''
mass-cut ($M_{\rm cut}(i)$) and the top of the incomplete Si-burning
region at $M_r = M_{\rm Si}$.  Then [Zn/Fe] in the mixed region
becomes as large as $\sim$ 0.5.

(2) Afterwards the mixed materials below $M_{\rm cut}(f)$ ($> M_{\rm
cut}(i)$) fall-back onto the compact remnant, and $M_{\rm cut}(f)$
becomes the final mass-cut. Then $M(^{56}$Ni) becomes smaller while
the mass ratios (Zn, Co, Mn)/Fe remain the same compared with the
values determined by $M_{\rm cut}(i)$.

We note that the occurrence of the mixing has been demonstrated by the
multi-D simulations of SN1987A and SNe Ib (e.g., Arnett et al. 1989;
Hachisu et al. 1990, 1991; Kifonidis et al. 2000), but the fall-back
simulations has been done only in 1D (Woosley \& Weaver 1995). 
Therefore, we need
multi-D simulations of fall-back to confirm the occurrence of the
``mixing and fall-back'' process and the resulting modification of 
the ejecta composition, which has not been done. Only when the mixing
takes place across the ``final mass-cut'', the SN yields are modified
by the mixing, which has not been taken into account in previous SN
yields.

\subsection{Aspherical Explosions}

\begin{table}
\caption{Models and Results of jet induced explosion models (Maeda \& Nomoto 2003). 
Masses are in solar mass unit (\Msun), 
and $\theta_{\rm jet}$ is in degree.}

\begin{center}
\begin{tabular}{cccccccccc}
\hline \hline
{Model} & {$M_{\rm MS}$} &  {$M_{\rm REM0}$} & {$\alpha$} &
{$\theta_{\rm jet}$} & {$E_{\rm total}$} &
{$M_{\rm REM}$} & {$M$($^{56}$Ni)} & [S/Si] & [C/O] \\
\hline
40A & 40 & 1.5 & 0.01 & 15 & 10.9 & 5.9 & 1.07E-1 & -0.46 & -1.3\\
40B & 40 & 1.5 & 0.01 & 45 & 1.2 & 6.8 & 8.11E-2  & -0.54 & -1.2\\
40C & 40 & 1.5 & 0.05 & 15 & 32.4 & 2.9 & 2.40E-1 & -0.30 & -1.3\\
40D & 40 & 3.0 & 0.01 & 15 & 8.5 &10.5 & 6.28E-8  & -1.1  & -1.0\\
\hline
25A & 25 & 1.0 & 0.01 & 15 & 6.7 & 1.9 & 7.81E-2  & -0.28 & -0.80\\
25B & 25 & 1.0 & 0.01 & 45 & 0.6 & 1.5 & 1.51E-1  & -0.26 & -0.82\\
\hline
\end{tabular}
\end{center}
\label{tab:2dtab}
\end{table}

The ``mixing and fall-back'' effect may also be effectively realized
in non-spherical explosions accompanying energetic jets (e.g., Maeda
\& Nomoto 2003; Maeda et al. 2002; Khokhlov et al. 1999; Nagataki et
al. 1997).  Compared with the spherical model with the same $M_{\rm
cut}(i)$ and $E$, the shock is stronger (weaker) and thus temperatures
are higher (lower) in the jet (equatorial) direction.  As a result, a
larger amount of complete Si-burning products are ejected in the jet
direction, while only incomplete Si-burning products are ejected in
the equatorial direction.  In total, complete Si-burning elements can
be enhanced (Maeda \& Nomoto 2003; Nomoto et al. 2001ab).

Recently, Maeda \& Nomoto (2003) have reported hydrodynamics and
nucleosynthesis of jet-induced explosion models. We show some of the
results below (See Maeda \& Nomoto 2003 for detailes). .

The main ingredient of the above models is a pair of jets propagating
through a stellar mantle. At the beginning of each calculation, the
central part ($M_r \leq M_{\rm REM0}$) of a progenitor star is
displaced by a compact remnant. The jets are injected at the interface
with the opening half-angle $\theta_{\rm jet}$. The energy injected by
the jets is assumed to be proportional to the accretion rate (i.e.,
$\dot E_{\rm jet} = \alpha \dot M_{\rm accretion} c^2$). The accretion
at the inner boundary is traced and is used to update the central
remnant' mass and the property of the jets. The models are summarized
in Table~\ref{tab:2dtab}.  The outcome of the explosion depends on the interaction
between the jets and the stellar mantle, and on the accretion rate
which is affected by the interaction itself. The
hydrodynamic evolution and nucleosynthesis in two dimensions are 
found to be as follows.

\begin{figure}[t]
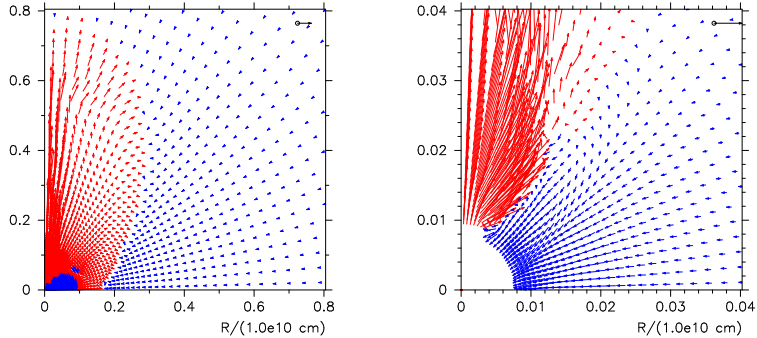

	\begin{center}
		\begin{minipage}[t]{0.45\textwidth}
			\includegraphics[width=0.925\textwidth]{f15a.ps}
		\end{minipage}
		\begin{minipage}[t]{0.45\textwidth}
			\includegraphics[width=0.925\textwidth]{f15b.ps}
		\end{minipage}
	\end{center}
\caption{Velocity distribution of Model 40A at 1.5 second after the 
initiation of the jets. 
The right panel shows that in the central region on an expanded scale. 
The reference arrow at the upper right represents $2 \times 10^9$cm s$^{-1}$. 
\label{fig:maedafig1}}
\end{figure}

Figure \ref{fig:maedafig1} shows a snapshot of the velocity distribution 
of Model 40A 
at 1.5 seconds after the initiation of the jets. 
The jets propagate through the stellar mantle, depositing their energies 
into ambient matter at the working surface. 
The bow shock expands laterally to push the stellar mantle sideways, 
which reduces the accretion rate. 
The strong outflow occurs along the $z$-axis (the jet direction), 
while matter accretes from the side. 
As the accretion rate decreases, the jets are turned off. 
Then the inflow along the $r$-axis turns to the weak outflow.

The outcome is a highly aspherical explosion. 
The accretion forms a central dense core. 
Densities near the center become much higher than those in spherical models. 
This feature is consistent with the suggestions by the spectroscopic 
(Mazzali et al. 2000) 
and the light curve (Nakamura et al. 2001a; Maeda et al. 2003) 
modeling of hypernovae. 

Other hydrodynamic properties are as follows (Tab.~\ref{tab:2dtab}): 

(1) A more massive star makes a more energetic explosion. 
The reason is that a more massive star has a stronger gravity to 
make the accretion rate higher. 
This is consistent with the relation seen in the hypernova branch 
(Fig. \ref{fig:nimass}).

(2) A more massive star forms a more massive compact remnant. The
remnant's mass increases as the accretion feeds it. The final mass
$M_{\rm REM}$ reaches typically $>5$\Msun \ for a $40$\Msun \ star, and
$\sim 2$\Msun \ for a $20$\Msun \ star. The bipolar models provide the way
of explosions with black hole formation in a consistent manner. Given
the discovery of the evidence of a hypernova explosion that
accompanied formation of a black hole of $\sim 5$\Msun \ (X-ray Nova
Sco; Israelian et al. 1999; Podsiadlowski et al. 2002), it offers an
interesting possibility.

\begin{figure}[t]
	\begin{center}
		\hspace{0.5cm}
		\begin{minipage}[t]{0.4\textwidth}
			\includegraphics[width=1.0\textwidth]{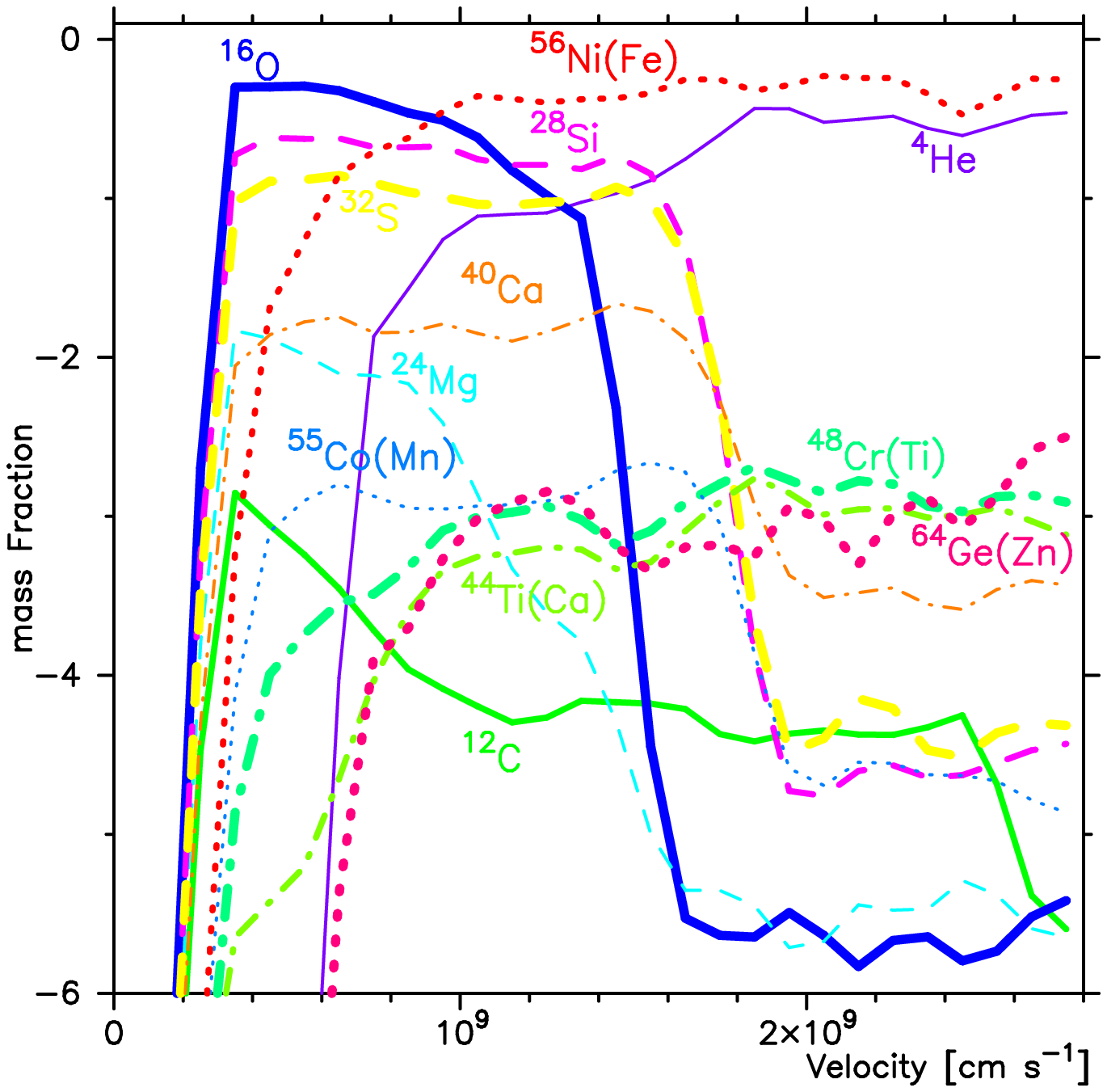}
		\end{minipage}
		\hspace{1cm}
		\begin{minipage}[t]{0.4\textwidth}
			\includegraphics[width=1.0\textwidth]{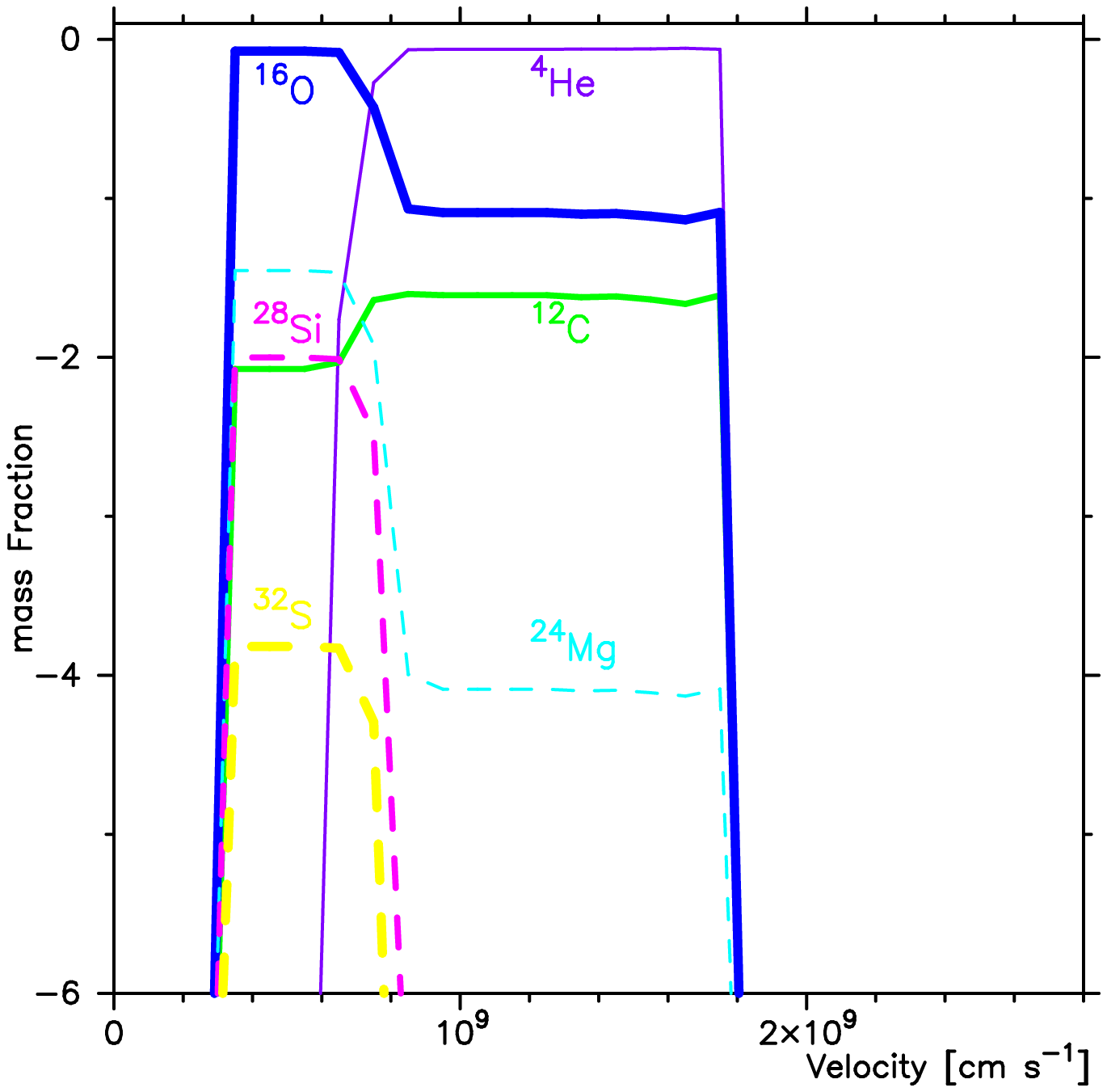}
		\end{minipage}
	\end{center}
	\caption{Mass fractions of selected isotopes in the velocity space along 
the $z$-axis (left panel) and along the $r$-axis (right panel) of Model 40A 
(Maeda \& Nomoto 2003). 
\label{fig:maedafig2}}
\end{figure}

Relatively high temperatures along the $z$-axis and low temperatures
along the $r$-axis have significant effects on nucleosynthesis. It
results in angle-dependent distribution of nucleosynthetic products as
shown in Figure \ref{fig:maedafig2}. The distribution of $^{56}$Ni (which
decays into $^{56}$Fe) is elongated along the $z$-axis, while that of
$^{16}$O is concentrated in the central region. Though this feature
has been already shown in Maeda et al. (2002) (see Fig.~\ref{eps10}), 
Maeda \& Nomoto (2003) have found that the central concentration 
of $^{16}$O and the
enhancement of the density there are much more significant than the
previous model of Maeda et al. (2002), because of the continuous
accretion from the side. Such a configuration, i.e., high velocity Fe
and low velocity O, has been suggested to be responsible for the
feature in the late phase spectra of SN1998bw, where the OI] 6300 was
narrower than the FeII] 5200 blend (Mazzali et al. 2001; Maeda et
al. 2002)

Along the $z$-axis, heavy isotopes which are produced with high
temperatures $T_9 \equiv T/10^9$K $\gsim 5$ are blown up to the
surface.  As a result, $^{64}$Ge, $^{59}$Cu, $^{56}$Ni, $^{48}$Cr, and
$^{44}$Ti (which decay into $^{64}$Zn, $^{59}$Co, $^{56}$Fe,
$^{48}$Ti, and $^{44}$Ca, respectively) are ejected at the highest
velocities. Isotopes which are synthesized with somewhat lower
temperatures ($T_9 = 4 - 5$) are first pushed aside as the jets
propagate, then experience circulation to flow into behind the working
surface. $^{55}$Co, $^{52}$Fe (which decay into $^{55}$Mn and
$^{52}$Cr, respectively), $^{40}$Ca, $^{32}$S, and $^{28}$Si are
therefore ejected at the intermediate velocities. Isotopes which are
not synthesized but are only consumed during the explosion are
accreted from the side. $^{24}$Mg, $^{16}$O, and $^{12}$C occupy the
innermost region at the lowest velocities.

\begin{figure}[t]
	\begin{center}
		\begin{minipage}[t]{0.45\textwidth}
			\includegraphics[width=1.\textwidth]{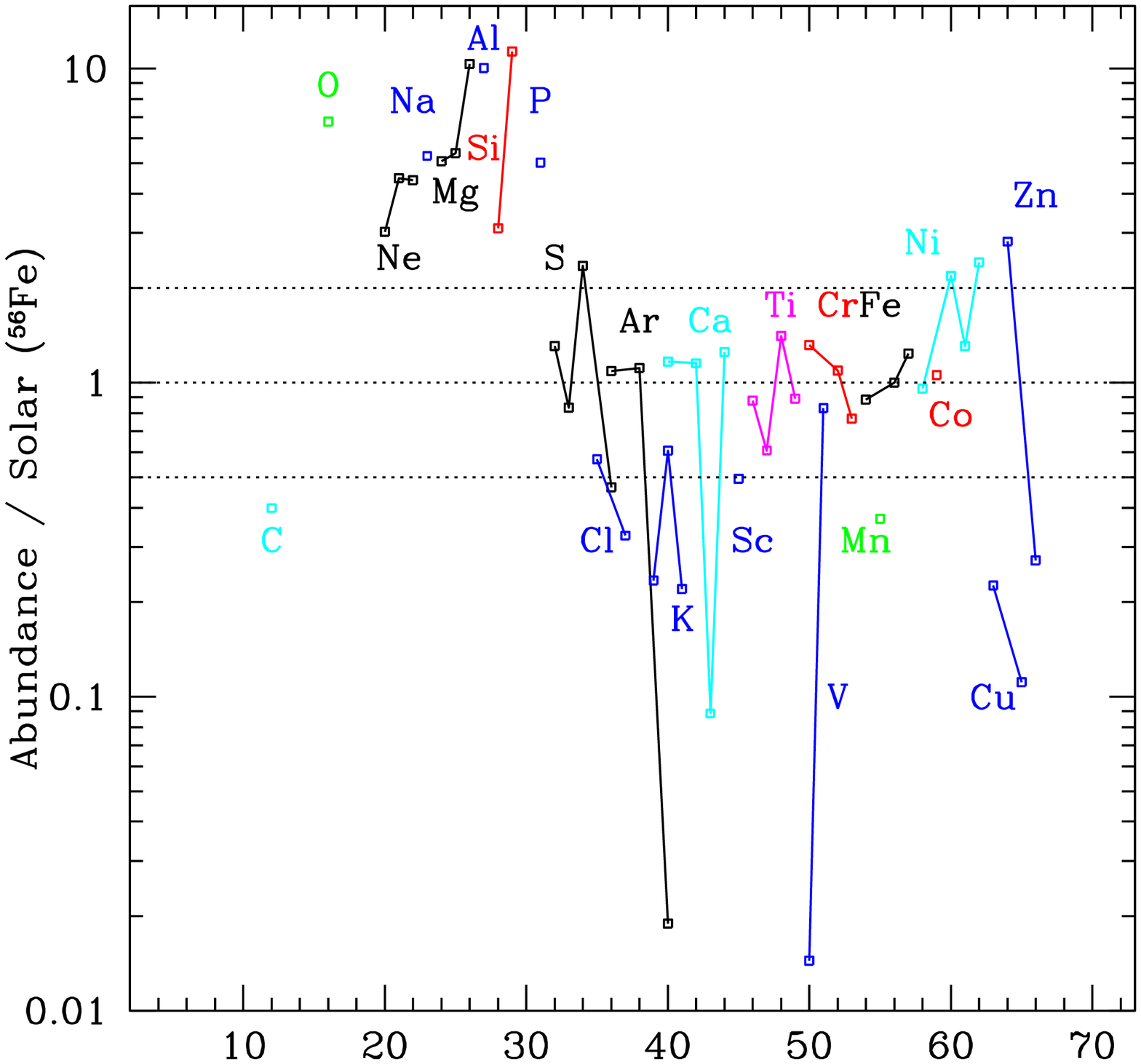}
		\end{minipage}
		\begin{minipage}[t]{0.45\textwidth}
			\includegraphics[width=1.\textwidth]{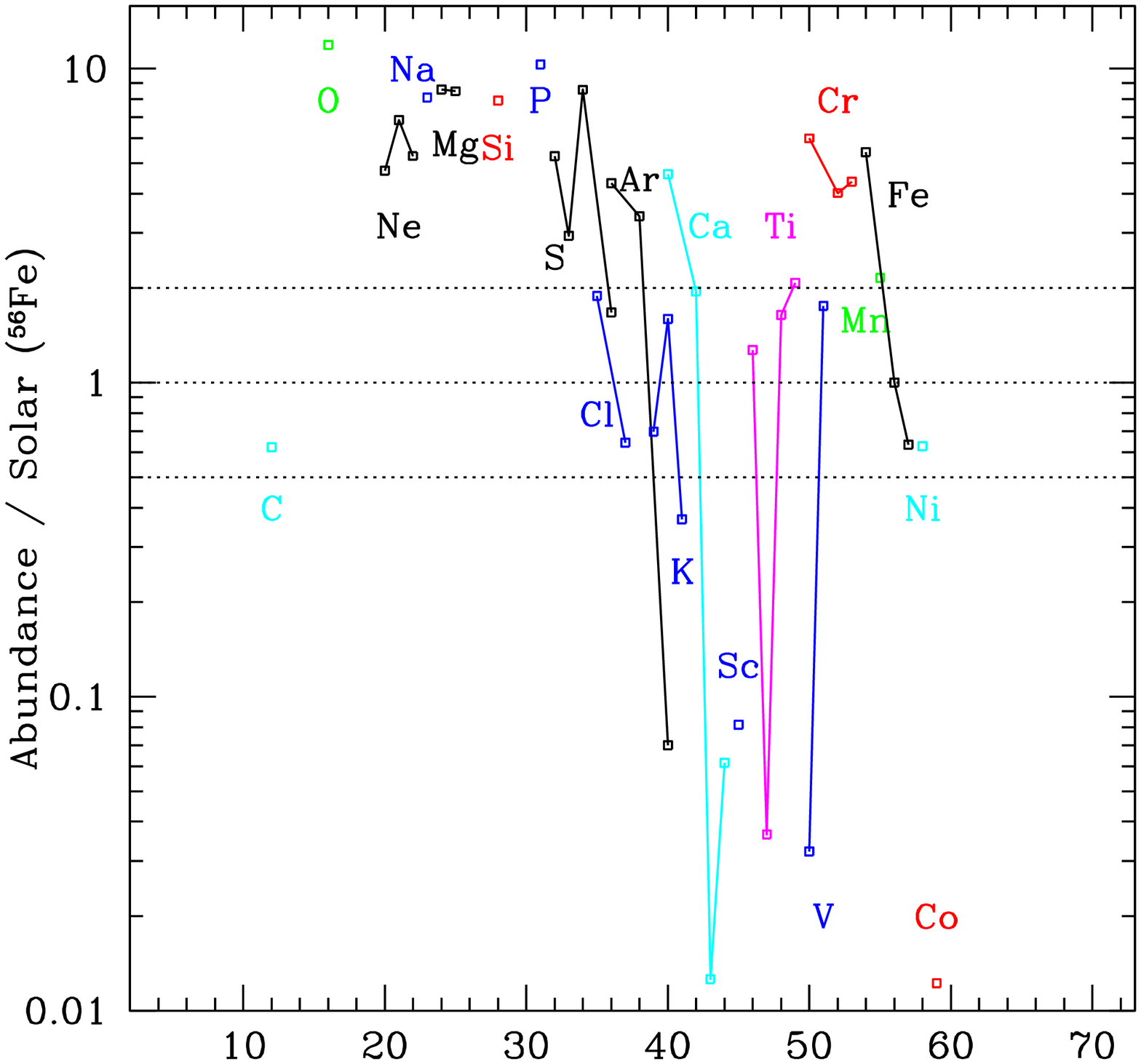}
		\end{minipage}
	\end{center}
\caption{Isotropic yields of the bipolar model 40A (left) and 
a spherical model (right) 
with $E_{51} = 10$, $M$($^{56}$Ni) $= 0.1$\Msun, and $M_{\rm MS}=40$\Msun \ 
(Maeda \& Nomoto 2003). 
\label{fig:maedafig3}}
\end{figure}

The distribution of isotopes as a function of the velocity shows
inversion as compared with conventional spherical models. This affects
the overall abundance patterns in the whole ejecta as shown in Figure
\ref{fig:maedafig3}. As noted above, materials which experience higher
$T_9$ are preferentially ejected along the $z$-axis, while materials
with lower $T_9$ accrete from the side in the bipolar models. Zn and
Co are ejected at higher velocities than Mn and Cr, so that the latter
accrete onto the central remnant more easily. As a consequence,
[Zn/Fe] and [Co/Fe] are enhanced, while [Mn/Fe] and [Cr/Fe] are
suppressed.

Interestingly, the abundance pattaerns of [(Zn, Co, Mn, Cr)/Fe] are
the same as seen in extremely metal-poor stars. The bipoar models have
the effect of ``mixing and fallback'' which has been suggested by
Umeda \& Nomoto (2002, 2003).

\section{Extremely Metal-Poor (EMP) Stars and Faint Supernovae}

Recently the most Fe deficient and C-rich low mass star, HE0107-5240,
was discovered (Christlieb et al. 2002).  This star has [Fe/H] $= -
5.3$ but its mass is as low as 0.8 \ms.  This would challenge the
recent theoretical arguments that the formation of low mass stars,
which should survive until today, is suppressed below [Fe/H] $= -4$
(Schneider et al. 2002).

The important clue to this problem is the observed abundance pattern
of this star.  This star is characterized by a very large ratios of
[C/Fe] = 4.0 and [N/Fe] = 2.3, while the abundances of elements
heavier than Mg are as low as Fe (Christlieb et al. 2002).
Interestingly, this is not the only extremely metal poor (EMP) stars
that have the large C/Fe and N/Fe ratios, but several other such stars
have been discovered (Ryan 2002).  Therefore the reasonable
explanation of the abundance pattern should explain other EMP stars as
well.  We show that the abundance pattern of C-rich EMP stars can be
reasonably explained by the nucleosynthesis of 20 - 130 \ms\
supernovae with various explosion energies and the degree of mixing
and fallback of the ejecta.

\begin{figure}[t]
  \begin{center}
  \begin{minipage}[t]{0.45\textwidth}
		\includegraphics[width=1.\textwidth]{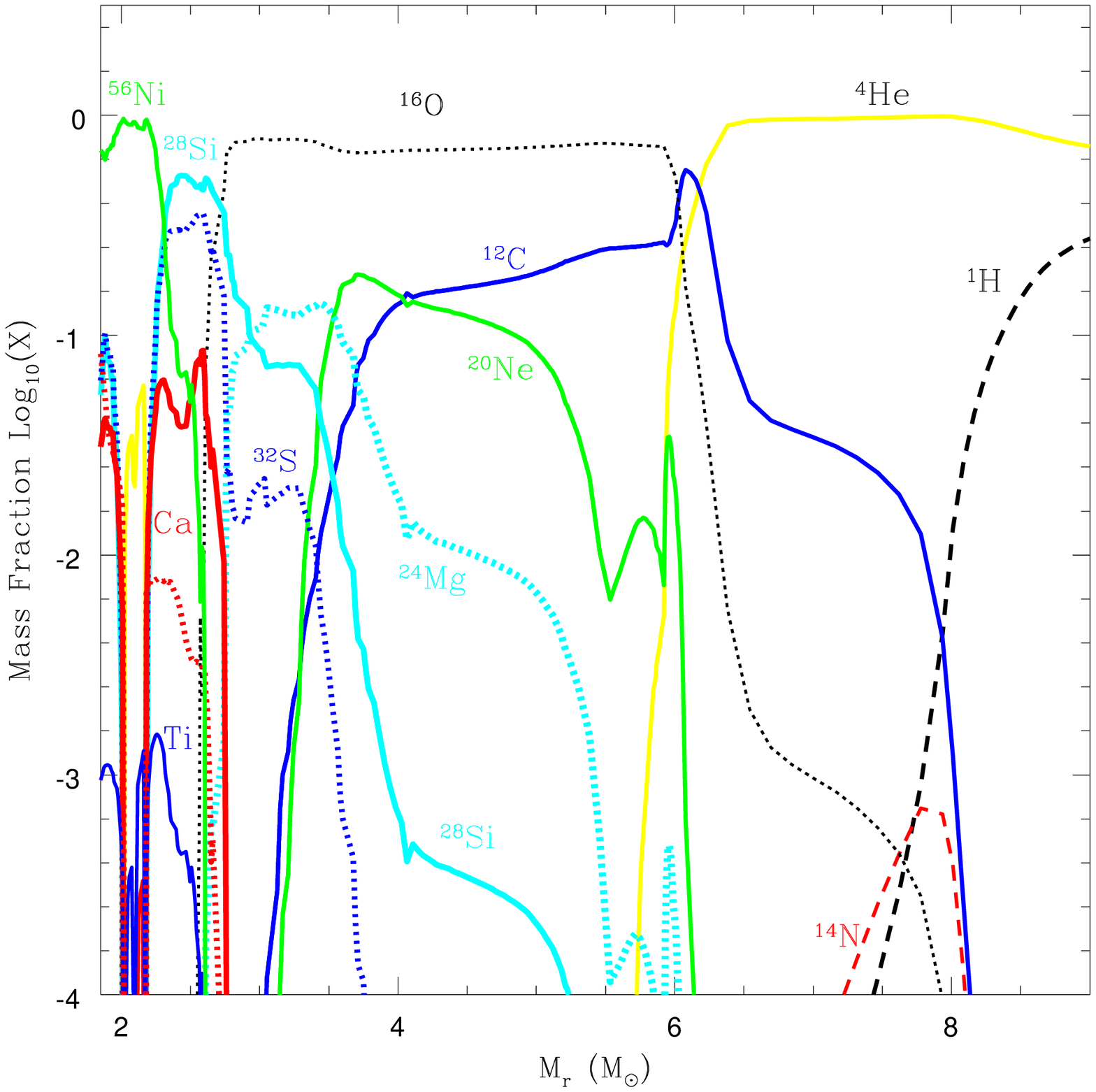}
  \end{minipage}
  \begin{minipage}[t]{0.49\textwidth}
		\includegraphics[width=1.\textwidth]{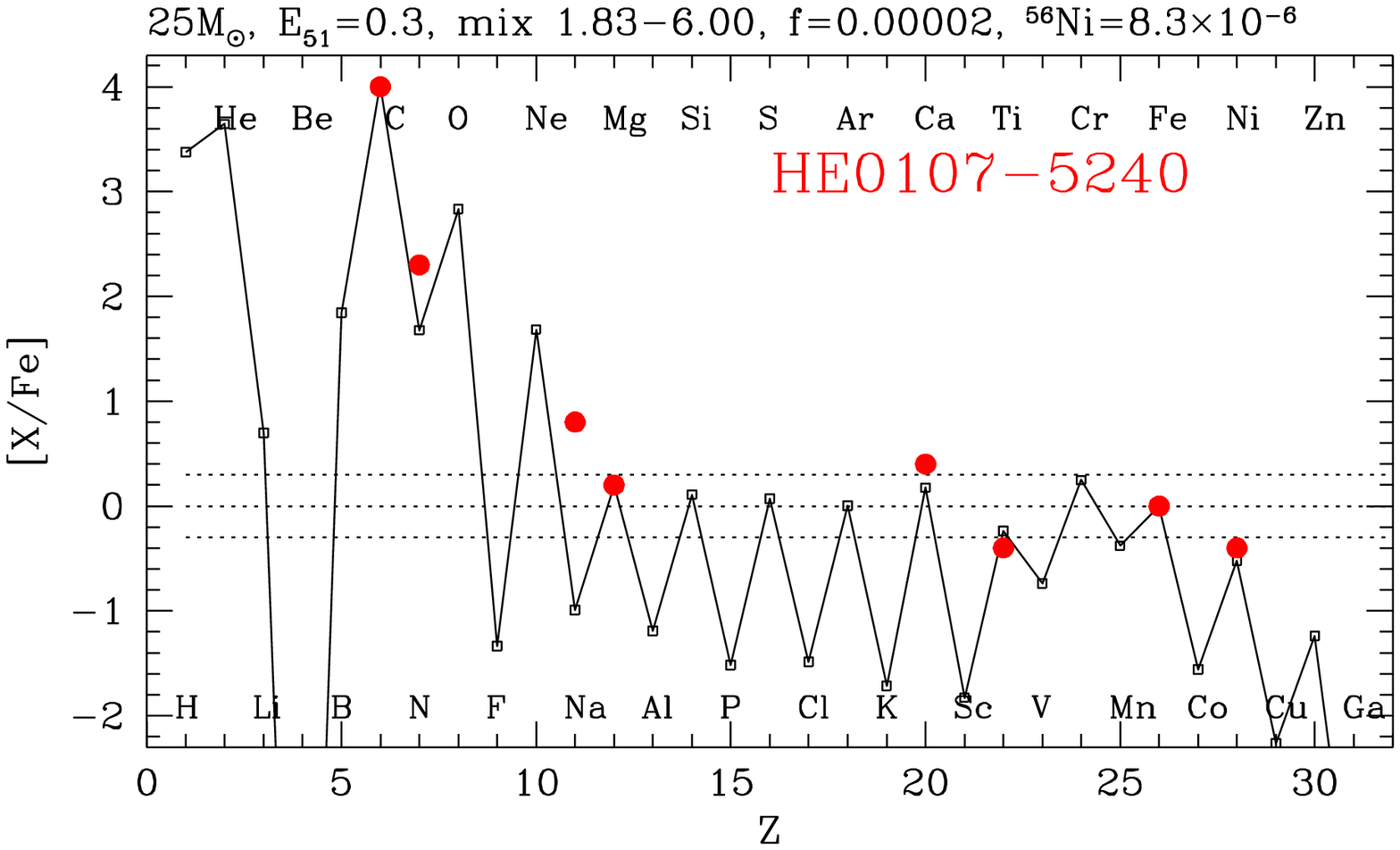}
  \end{minipage}
 \end{center}
\caption{
(left) The post-explosion abundance distributions for the 25 $M_\odot$
model with the explosion energy $E_{51} =$ 0.3 (right).
Elemental abundances of the C-rich most Fe deficient star HE0107-5240
(filled circles), compared with a theoretical supernova yield
(Umeda \& Nomoto 2003).
\label{fig:25m}}
\end{figure}

\subsection{The Most Metal-Poor Star HE0107-5240}

We consider a model that C-rich EMP stars are produced in the ejecta
of (almost) metal-free supernova mixed with extremely metal-poor
interstellar matter.  We use Pop III pre-supernova progenitor models,
simulate the supernova explosion and calculate detailed
nucleosynthesis (Umeda \& Nomoto 2003).

In Figure \ref{fig:25m} (right) we show that the elemental abundances
of one of our models are in good agreement with HE0107-5240, where the
progenitor mass is 25 \ms\ and the explosion energy $E_{51} =$ 0.3
(Umeda \& Nomoto 2003).

In this model, explosive nucleosynthesis takes place behind the shock
wave that is generated at $M_r =$ 1.8 \ms\ and propagates outward. The
resultant abundance distribution is seen in Figure \ref{fig:25m}
(left), where $M_r$ denotes the Lagrangian mass coordinate measured
from the center of the pre-supernova model (Umeda \& Nomoto 2003).
The processed material is assumed to mix uniformly in the region from
$M_r =$ 1.8 \ms\ and 6.0 \ms.  Such a large scale mixing was found to
take place in SN1987A and various explosion models (e.g., 
Hachisu et al. 1990;
Kifonidis et al. 2000).  Almost all materials below $M_r =$ 6.0 \ms\
fall back to the central remnant and only a small fraction ($f = 2
\times$ 10$^{-5}$) is ejected from this region.  The ejected Fe mass
is 8 $\times$ 10$^{-6}$ \ms.

The CNO elements in the ejecta were produced by pre-collapse He shell
burning in the He-layer, which contains 0.2 \ms\ $^{12}$C.  Mixing of
H into the He shell-burning region produces 4 $\times$ 10$^{-4}$ \ms\
$^{14}$N.  On the other hand, only a small amount of heavier elements
(Mg, Ca, and Fe-peak elements) are ejected and their abundance ratios
are the average in the region of $M_r =$ 1.8 - 6.0 \ms. The sub-solar
ratios of [Ti/Fe] $= -0.4$ and [Ni/Fe] $= -0.4$ are the results of the
relatively small explosion energy ($E_{51} =$ 0.3).  With this "mixing
and fallback", the large C/Fe and C/Mg ratios observed in HE0107-5240
are well reproduced (Umeda \& Nomoto 2003).

In this model, N/Fe appears to be underproduced. However, N can be
produced inside the EMP stars through the C-N cycle, and brought up to
the surface during the first dredge up stage while becoming a
red-giant star (Boothroyd \& Sackmann 1999).

\subsection{Carbon-rich EMP stars: CS 22949-037 and CS 29498-043} 

\begin{figure}[t]
  \begin{center}
  \begin{minipage}[t]{0.49\textwidth}
		\includegraphics[width=1.\textwidth]{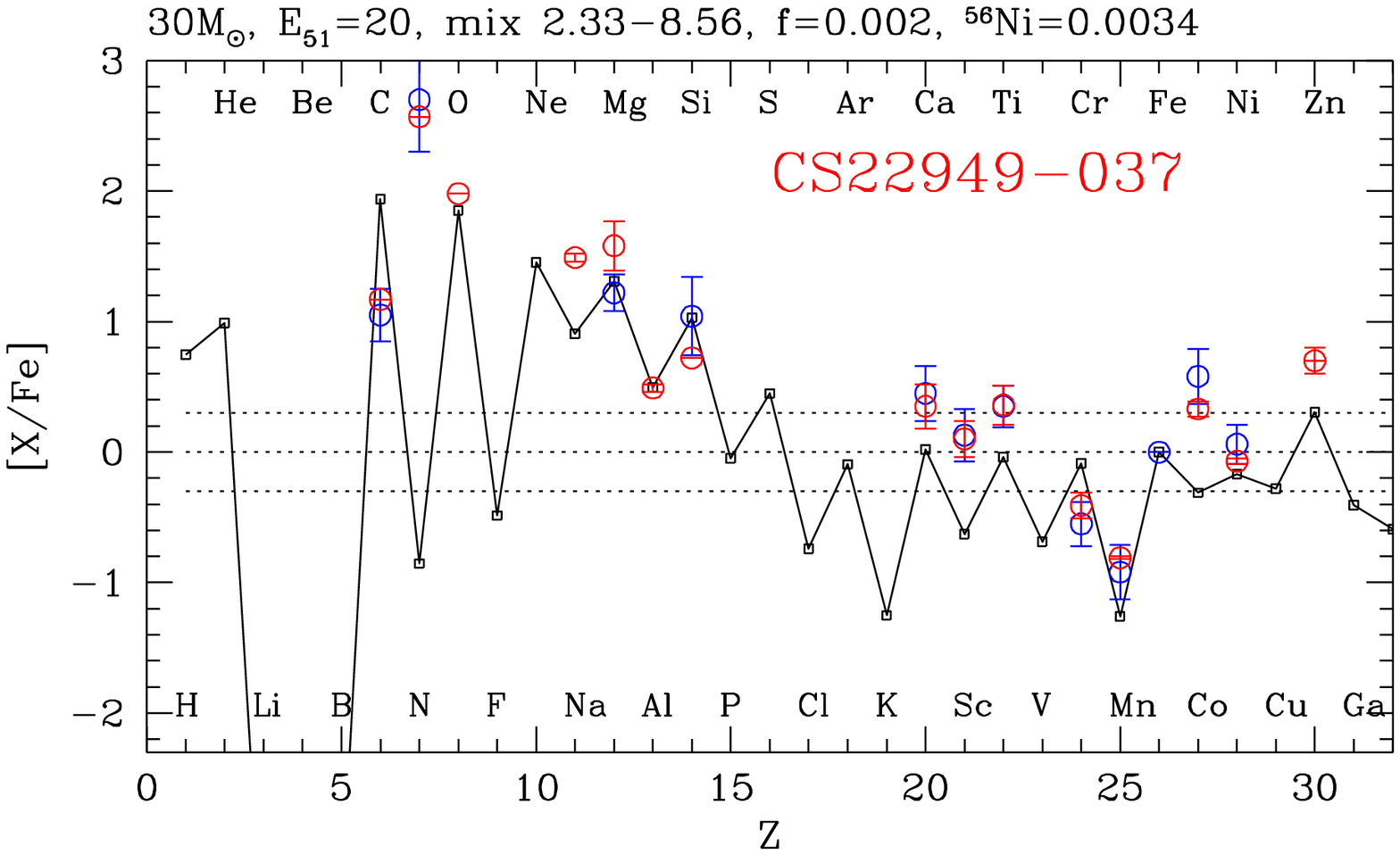}
  \end{minipage}
  \begin{minipage}[t]{0.49\textwidth}
		\includegraphics[width=1.\textwidth]{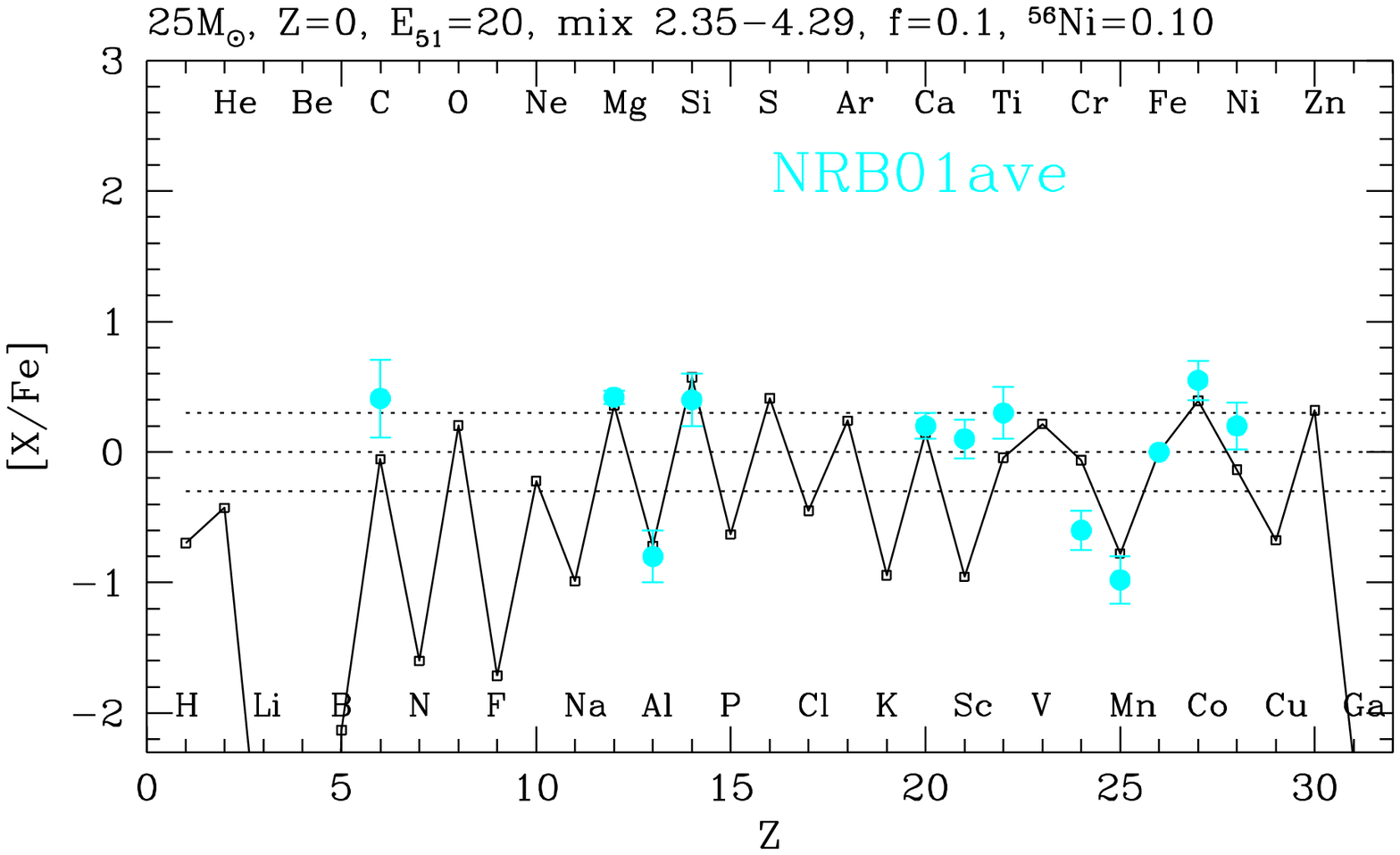}
  \end{minipage}  \end{center}
\caption{
(left) Elemental abundances of CS 22949-037 (open circles for Norris et al. (2001), 
and solid squares for Depagne et al. (2002)), compared
with a theoretical supernova yield (Umeda \& Nomoto 2003). 
(right) Averaged
elemental abundances of stars with [Fe/H] $= -3.7$ (Norris et al. 2001).
\label{fig:emp2}}
\end{figure}

The "mixing and fallback" is commonly required to reproduce the
abundance pattern of typical EMP stars.  In Figure \ref{fig:emp2}
(left) we show a model, which is in good agreement with CS22949-037
(Umeda \& Nomoto 2003).  This star has [Fe/H] $= -4.0$ and also C,
N-rich (Norris et al. 2001; Depagne et al. 2002), though C/Fe and N/Fe
ratios are smaller than HE0107-5240.  The model is the explosion of a
30 \ms\ star with $E_{51} =$ 20.  In this model, the mixing region
($M_r =$ 2.33 - 8.56 \ms) is chosen to be smaller than the entire He
core ($M_r =$ 13.1 \ms) in order to reproduce relatively large Mg/Fe
and Si/Fe ratios.  Similar degree of the mixing would also reproduce
the abundances of CS29498-043 (Aoki et al. 2002), which shows similar
abundance pattern.

We assume a larger fraction of ejection, 2\%, from the mixed region
for CS22949-037 than HE0107-5240, because the C/Fe and N/Fe ratios are
smaller. The ejected Fe mass is 0.003 \ms. The larger explosion energy
model is favored for explaining the large Zn/Fe, Co/Fe and Ti/Fe
ratios (Umeda \& Nomoto 2002).

Without mixing, elements produced in the deep explosive burning
regions, such as Zn, Co, and Ti, would be underproduced. Without
fallback the abundance ratios of heavy elements to lighter elements,
such as Fe/C, Fe/O, and Mg/C would be too large.  In this model, Ti,
Co, Zn and Sc are still under-produced. However, these elements may be
enhanced efficiently in the aspherical explosions (Maeda et al. 2002;
Maeda \& Nomoto 2003).  Almost the same effects as the "mixing and
fallback mechanism" are realized if the explosion is jet-like,
although the total energy can be smaller if the beaming angle of the
jet is small (Maeda \& Nomoto 2003).

According to Maeda \& Nomoto (2003), some bipolar models explain the
existence of Fe-poor explosions with very little amount of Fe, which
leave a massive central remnant with $M_{\rm REM} \sim 10$\Msun. Such
explosions would be responsible for the formation of the carbon-rich
metal-poor stars, e.g., CS22949-037 (Norris et al. 2001; Aoki et
al. 2002; Depagne et al. 2002), as the bipoplar explosions with
smaller $M$($^{56}$Ni) lead to larger [C/O] (See also Umeda \& Nomoto
2003).

\subsection{EMP Stars with a Typical Abundance Pattern}

Similarly, the "mixing and fall back" process can reproduce the
abundance pattern of the typical EMP stars without enhancement of C
and N.  Figure \ref{fig:emp2} (right) shows that the averaged
abundances of [Fe/H] $= -3.7$ stars in Norris et al. (2001) can be
fitted well with the model of 25 \ms\ and $E_{51} =$ 20 but larger
fraction ($\sim$ 10\%) of the processed materials in the ejecta.

\subsection{SNe with Small Fe Ejection}

In our model, [Fe/H] of several kinds of EMP stars can be understood
in the supernova-induced star formation scheme (Audouze \& Silk 1995;
Shigeyama \& Tsujimoto 1998; Nakamura et al. 1999). In this scheme,
[Fe/H] of the second-generation stars are determined by the ejected Fe
mass divided by the mass of hydrogen swept up by the supernova
ejecta. As the swept-up hydrogen mass is roughly proportional to the
explosion energy, Fe/H $\propto$ (M(Fe)/0.07$M_\odot$) / E$_{51}$,
where M(Fe) is the ejected Fe mass. The average stars of [Fe/H]
$\simeq -3.7$ (Norris et al. 2001), CS22949-037, and HE0107-5240
correspond to (M(Fe)/0.07$M_\odot$) / E$_{51}$ = 0.07, 0.002, and
0.0004, respectively. This correspondence suggests that [Fe/H] of the
EMP stars do not reflect the age of the stars, but rather the
properties of the supernovae, such as the degree of mixing and
fallback or collimation of a jet.

We have also shown that the most iron-poor star, as well as other
C-rich EMP stars, is likely to be enriched by massive supernovae that
are characterized by relatively small Fe ejection. Such supernovae are
not hypothetical, but have been actually observed, forming a distinct
class of type II supernovae (`faint supernovae') (Nomoto et
al. 2002). The prototype is SN1997D, which is very underluminous and
shows quite narrow spectral lines (Turatto et al. 2000): these
features are well modeled as an explosion of a 25$M_\odot$ star
ejecting only $2 \times 10^{-3} M_\odot$ $^{56}$Ni with small
explosion energy E$_{51} \sim 0.4$ (Turatto et al. 2000). SN1999br is
a very similar faint supernova (Zampieri et al. 2003). On the other hand,
typical EMP stars without enhancement of C and N correspond to the
abundance pattern of energetic supernovae (`hypernovae'). For
both cases, black holes more massive than$\sim 3 - 10M_\odot$ must be
left as results of fallback, suggesting copious formation of the first
black holes from the first stars. These black holes may make up some
of the dark mass in the Galactic halo.

In our model, HE0107-5240 with [Fe/H]= $-5.3$ was formed from C- and
O-enhanced gases with [(C,O)/H] $\sim -1$. With such enhanced C and O,
the cooling efficiency is large enough to form small-mass stars. In
other word, our model predicts that low-mass EMP stars with [Fe/H] $ <
-4$ are likely to have enhanced [(C, N, O)/Fe] and [Mg/Fe] in some
cases.  A consequence of the low-mass EMP stars being carbon-rich is
that the population III stars that provided their metals are massive
enough to form (the first) black holes.

\section{The First Stars}

It is of vital importance in current astronomy to identify the first
generation stars in the Universe, i.e., totally metal-free, Pop III
stars.  The impact of the formation of Pop III stars on the evolution
of the Universe depends on their typical masses.  Recent numerical
models have shown that, the first stars are as massive as $\sim$ 100
\ms\ (Abel, Bryan, \& Norman 2002).  The formation of long-lived low
mass Pop III stars may be inefficient because of slow cooling of metal
free gas cloud, which is consistent with the failure of attempts to
find Pop III stars.

If the most Fe deficient star, HE0107-5240, is a Pop III low mass star
that has gained its metal from a companion star or interstellar matter
(Yoshii 1981), would it mean that the above theoretical arguments are
incorrect and that such low mass Pop III stars have not been
discovered only because of the difficulty in the observations?

Based on the results in the earlier section, we propose that the first
generation supernovae were the explosion of $\sim$ 20-130 \ms\ stars
and some of them produced C-rich, Fe-poor ejecta.  Then the low mass
star with even [Fe/H] $< -5$ can form from the gases of mixture of
such a supernova ejecta and the (almost) metal-free interstellar
matter, because the gases can be efficiently cooled by enhanced C and
O ([C/H] $\sim -1$).

\begin{table}[t]
\caption{The results of the stability analysis for Pop III and Pop I
stars.  $\bigcirc$ and $\times$ represent that the star is stable and
unstable, respectively.  The $e$-folding time for the fundamental mode
is shown after $\times$ in units of $10^4$yr (Nomoto et al. 2003).}
\begin{center}
\footnotesize
\begin{tabular}{ccccccc}
\hline \hline
{\large mass ($M_\odot$)} &{\large 80}&{\large 100}&{\large 
120}&{\large 150} &{\large 180} &{\large 300} \\ \hline
{\large Pop III} &{\large $\bigcirc$ }&{\large $\bigcirc$ }&{\large
$\bigcirc$ }&{\large $\times$ (9.03)} &{\large 
 $\times$ (4.83)} &{\large $\times$ (2.15)} \\ 
{\large Pop I }&{\large $\bigcirc$ }&{\large $\times$ (7.02)} &{\large $\times$ (2.35)} &{\large $\times$ (1.43)} &{\large
 $\times$ (1.21)} &{\large $\times$ (1.71)} \\ \hline 
\end{tabular}
\end{center}
\label{tab:pop3}
\end{table}

\subsection {Pair-Instability Supernovae}

We have shown that the ejecta of core-collapse supernova explosions of
massive stars can well account for the abundances of EMP stars. We can
put further constraint on the typical mass of Pop III. The abundances
of all observed EMP stars including the most metal-poor one are
inconsistent with the abundance pattern of pair-instability supernovae
(PISNe). For typical EMP stars and CS22948-037, enrichment by PISNe
cannot be consistent with the observed abundant Zn/Fe and Co/Fe ratios
(Umeda \& Nomoto 2002; Heger \& Woosley 2002). For HE0107-5240 and
other C-rich EMP stars, PISNe enrichment is difficult to account for
the large C/Mg ratios. Therefore the supernova progenitors that are
responsible for the formation of EMP stars are in the range of $M \sim
20 - 130$ \ms, but not more massive than 130 \ms.  This upper limit
depends on the stability of massive stars as will be discussed below.

\subsection{Stability and Mass Loss of Massive Pop III Stars}

To determine the upper limit mass of the Zero Age Main Sequence
(ZAMS), we analyze a linear non-adiabatic stability of massive
($80M_{\odot}$ - $300M_{\odot}$) Pop III stars using a radial
pulsation code (Nomoto et al. 2003).  Because CNO elements are absent
during the early stage of their evolution, the CNO cycle does not
operate and the star contracts until temperature rises sufficiently
high for the $3\alpha$ reaction to produce $^{12}$C.  We calculate
that these stars have $X_{\rm CNO} \sim 1.6 - 4.0\times10^{-10}$, and
the central temperature $T_{c}\sim1.4\times10^8 K$ on their ZAMS.  We
also examine the models of Pop I stars for comparison.

Table~\ref{tab:pop3} shows the results for our analysis. The critical mass of ZAMS
Pop III star is $128M_{\odot}$ while that of Pop I star is
$94M_{\odot}$.  This difference comes from very compact structures
(with high $T_c$) of Pop III stars.

Stars more massive than the critical mass will undergo pulsation and
mass loss. We note that the $e$-folding time of instability is much
longer for Pop III stars than Pop I stars with the same mass, and thus
the mass loss rate is much lower. These results are consistent with
Ibrahim, Boury, \& Noels (1981) and Baraffe, Heger, \& Woosley (2001).
However, the absence of the indication of PISNe may imply that these
massive stars above 130 \ms\ undergo significant mass loss, thus
evolving into Fe core-collapse rather than PISNe.

\section{Concluding Remarks}\label{sec:summary}

In this paper, we first describe how the basic parameters of hypernova
SN~1998bw are derived from observations and modeling, and discuss
the properties of other hypernovae individually.  These hypernovae
seem to come from rather massive stars, being more massive than $\sim$
20 - 25 \ms\ on the main-sequence, thus forming black holes.  On the
other hand, there are some examples of massive SNe with only a small
kinetic energy.  We suggest that stars with non-rotating black holes
are likely to collapse "quietly" ejecting a small amount of heavy
elements (Faint supernovae).  In contrast, stars with rotating black
holes are likely to give rise to very energetic supernovae
(Hypernovae).  

We present distinct nucleosynthesis features of these
two types of "black-hole-forming" supernovae.  Nucleosynthesis in
Hypernovae are characterized by larger abundance ratios
(Zn,Co,V,Ti)/Fe and smaller (Mn,Cr)/Fe.  Nucleosynthesis in Faint
supernovae is characterized by a large amount of fall-back, 
yielding large [$\alpha$/Fe]. 
We show
that the abundance pattern of the recently discovered most Fe
deficient star, HE0107-5240, and other extremely metal-poor
carbon-rich stars are in good accord with those of black-hole-forming
supernovae, but not pair-instability supernovae.  This suggests that
black-hole-forming supernovae made important contributions to the
early Galactic (and cosmic) chemical evolution.  Finally we discuss
the nature of First (Pop III) Stars.

\begin{acknowledgments}

This work has been supported in part by the grant-in-Aid for
Scientific Research (14047206, 14540223) of the Ministry of Education,
Science, Culture, and Sports in Japan.

\end{acknowledgments}

\end{document}